\documentclass[aps,superscriptaddress,twocolumn,longbibliography,floatfix]{revtex4-1}
\usepackage{graphicx} % Required for inserting images
\usepackage{amsmath}
\usepackage{amssymb}
\usepackage{amsfonts}
\usepackage{bbold}
\usepackage{bm}
\usepackage{color}
\usepackage{braket}
\usepackage[colorlinks,linkcolor=blue,citecolor=blue,allcolors=blue]{hyperref}

\begin{document}

\title{Curvature-induced nonlinear anomalous Hall effect in thin magnetic shells}

\author{Maria Teresa Mercaldo}
\affiliation{Dipartimento di Fisica ``E. R. Caianiello", Universit\`a di Salerno, IT-84084 Fisciano (SA), Italy}

\author{Mario Cuoco}
\affiliation{CNR-SPIN, c/o Universit\`a di Salerno, IT-84084 Fisciano (SA), Italy}

\author{Carmine Ortix}
\email{cortix@unisa.it}
\affiliation{Dipartimento di Fisica ``E. R. Caianiello", Universit\`a di Salerno, IT-84084 Fisciano (SA), Italy}
\affiliation{CNR-SPIN, c/o Universit\`a di Salerno, IT-84084 Fisciano (SA), Italy}

%\date{\today} 
\begin{abstract}
Optoelectronic and nonlinear transport experiments probe the quantum geometric tensor of Bloch states, whose real and imaginary components --the quantum metric and the Berry curvature-- are typically constrained by symmetry. Here, we show that geometric bending provides a route to engineer such responses in centrosymmetric ferromagnets. Curvature-induced strain gradients across the shell thickness break inversion symmetry and activate an orbital Rashba coupling. In the presence of in-plane magnetization and spin-orbit coupling, this generates spin textures with a nontrivial quantum geometry, leading to an intrinsic nonlinear anomalous Hall effect (NAHE) governed by the quantum metric and maximized when the magnetization aligns with the applied electric field. When geometric deformations further break  twofold rotational symmetry around the out-of-plane axis, an additional NAHE emerges, maximal for magnetization perpendicular to the driving electric field and governed by the Berry curvature dipole, thus giving access to the imaginary component of the quantum geometric tensor. These results establish curved ferromagnetic shells as a platform for engineering anisotropic nonlinear transport and for selectively probing both components of the quantum geometric tensor.
\end{abstract}

\maketitle

\section{Introduction}
It has been recently established that optoelectronic and  nonlinear transport experiments can give direct access to physical quantities which encode the geometric properties of the electronic Bloch waves. These are determined by a quantum geometric tensor (QGT) that measures the infinitesimal distance between Bloch states at different points of the Brillouin zone (BZ)~\cite{pro80}. The imaginary part of the QGT corresponds to the well-known Berry curvature (BC), which provides the Chern number classifying two-dimensional crystalline insulators~\cite{Thouless}. In metallic systems with broken time-reversal symmetry, instead, the Berry curvature yields a non-vanishing Berry phase that governs the intrinsic part of the anomalous Hall conductivity~\cite{Haldane2,nag10}. 
In non-magnetic materials where inversion symmetry is broken, local concentration of BC can be probed in a second-order Hall effect that can possess a contribution scaling linearly with the electronic relaxation time $\tau$. This contribution is induced by the first moment of the BC in momentum space -- the so-called Berry curvature dipole~\cite{Sodemann2015,Ortix2021,Du21,Suarez2025}. 
Although nonlinear transversal currents can generally exist in trigonal systems, including $(111)$ surfaces of centrosymmetric materials such as Bi$_2$Se$_3$~\cite{he21} and elemental Bi~\cite{Makushko24}, and two-dimensional materials as bilayer graphene~\cite{iso20}, the appearance of a BCD is restricted to materials with unusually low crystalline symmetries, particularly in two-dimensional electronic systems. For this material class, the largest symmetry symmetry group compatible with a BCD is ${\mathcal C}_s$ which contains a single vertical mirror line. Non-vanishing BCD have been predicted at the surfaces of SnTe~\cite{hsi12} in its ferroelectric phase~\cite{lau19}, in transition metal dichalcogenides~\cite{Ma2019,kan19}, in strained bilayer graphene~\cite{bat19,ho21} as well as at $(111)$ oxide heterointerfaces~\cite{les23}.

\begin{figure*}
    \centering
    \includegraphics[width=0.7\textwidth]{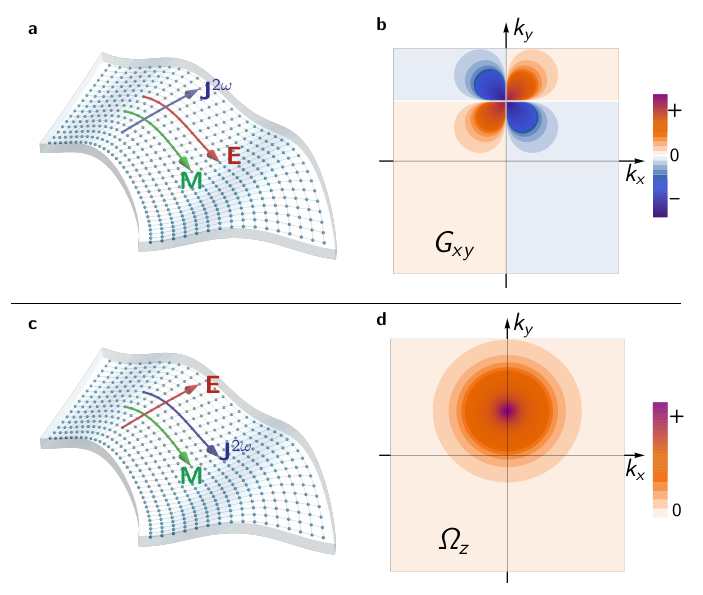}
\caption{ {\bf Curvature induced nonlinear  Hall effects.} {\bf a} Sketch of a curved ferromagnetic thin shell: the green and red arrows show the orientation of the magnetization $\mathbf{M}$ and of the driving electric field $\mathbf{E}$, respectively. When the two vectors are parallel a second-harmonic transverse current $J^{2\omega}$ is observed induced by the quantum metric -- the real part of the quantum geometric tensor (QGT) -- of the electronic wavefunctions. {\bf b} Momentum-space map of one of the three components of the band normalized quantum metric, namely $G_{xy}$, 
for a representative case. %physical case of curved ferromagnetic membrane. 
{\bf c} Sketch of a bent ferromagnetic membrane with applied electric field $\mathbf{E}$ (red arrow) orthogonal to the magnetization $\mathbf{M}$ (green arrow). In this situation the second-harmonic transverse current $J^{2\omega}$ is parallel to the magnetization. This response is activated by the Berry curvature -- the imaginary part of the QGT -- whose density plot in momentum space is drawn in {\bf d}.  
    \label{fig:fig1}}
\end{figure*}
The quantum metric of the electronic Bloch waves -- the real part of the QGT -- can also induce electrical currents quadratic in the driving electric field (both in the longitudinal and in the transversal channel) provided time-reversal symmetry is broken~\cite{gao14,das23,kap24,Jiang2025,tor23}. This contribution is independent of the electronic relaxation time and thus intrinsic. The observation of quantum metric-induced transport has been so far limited to the topological antiferromagnet~\cite{gao23,wan23exp} MnBi$_2$Te$_4$ and, as a nonlinear magnetoresistance, to LaAlO$_3$/SrTiO$_3$ heterointerfaces~\cite{Sala_24} and three-dimensional topological insulators~\cite{Mercaldo2025,Sala2026}. In this article, we prove that signatures of quantum metric can be observed in a large class of easy-surface ferromagnetic thin shell of centrosymmetric materials undergoing geometric bending. The strain field along the thickness of the curved shell is a source of inversion symmetry breaking that activates a linear spin-orbital coupling, known as orbital Rashba coupling~\cite{Park11,Kim13,Mer20}, and an ensuing Rashba spin-orbit coupling. The tendency towards spin-momentum locking imposed by Rashba spin-orbit coupling on the in-plane spin magnetization of the ferromagnet creates spin textures that are characterized by a non-trivial quantum geometric tensor. We will show that the corresponding band-energy normalized quantum metric~\cite{Mercaldo2025} possesses a non-vanishing net dipole that drives a nonlinear anomalous Hall current $j_a=\sigma_{a;bb} E_b^2$ with $E_b$ the driving electric field and ${\hat a} \perp {\hat b}$, that is maximal when the saturated magnetization is parallel to the driving electric field [see Fig.~\ref{fig:fig1}(a),(b)]. This intrinsic, quantum metric-induced, nonlinear anomalous Hall effect (NAHE) is accompanied by a nonlinear Drude contribution that can be parsed thanks to the different dependence on the relaxation time $\tau$. We will also show that geometric deformations that cause a loss of in-plane twofold rotational symmetry can drive an additional NAHE that is instead maximal when the in-plane magnetization is perpendicular to the driving electric field [see Fig.~\ref{fig:fig1}(c)]. This NAHE is induced by a curvature-induced BCD [see Fig.~\ref{fig:fig1}(d)] that is absent in the conventional flat material structure, and thus provides a footprint of the imaginary part of the QGT.

The curvature-induced NAHE we propose here represents a new physical effect in curvilinear magnetism. Curvilinear magnetism has recently emerged as a powerful approach to modify the magnetic state and thus the responses of magnetic materials. 
It has been originally conceived taking advantage of the coupling between the nanoscale shape of a magnetic structure and its magnetic order parameter~\cite{Streubel16,Sheka21b,Makarov22}. Generally speaking, the confinement of the magnetic energy functional to a bent two-dimensional surface is reflected in the appearance of an effective Dzyaloshinskii-Moriya interaction coupling and an effective anisotropy~\cite{Gaididei14,Kravchuk16}. The curvature-induced DMI has been shown, for instance, to stabilize modulated phases even in the complete absence of any materials' intrinsic DMI~\cite{Yershov2020}. 
Chiral responses stemming from the exchange-driven DMI have been experimentally verified~\cite{Volkov19c} using parabolic stripes of Ni$_{81}$Fe$_{19}$ (Py). 
%In curved stripes with curvature gradients, the curvature-induced DMI can drive domain wall motion in the complete absence of any external stimulus. 
%Curvilinear three-dimensional structures of helimagnetic materials have been also proposed as artificial magnetoelectric devices when embedded in a piezoelectric matrix: since the magnetic state can be controlled by changing the geometry of the curved helimagnetic wire, the piezoelectric matrix allows an electric field-induced switching between different magnetic states that can be assigned logical ``1" or ``0". 
%\textcolor{red}{comment magnetostatic terms.}
The exchange-induced anisotropy sets instead the magnetic ground state of systems with easy-surface type of magnetocrystalline anistropy without any reorganization of the magnetic ground state. For example in one-dimensional rippled magnetic nanomebranes the stable ground state has a magnetization that is simply oriented perpendicular to the ripples and thus along the translationally invariant direction~\cite{Ortix2023}. 
%\textcolor{red}{Check magnetic ground states in generic bent surface with easy-surface MCA}
Our study demonstrates that despite this absence of curvature effects on the magnetic ground state, the real space geometry can generate non-trivial quantum geometric properties and yield  intrinsic nonlinear transport phenomena.

\section{Curvature-induced orbital and spin Rashba coupling}
In this section we discuss how in quasi-two-dimensional (magnetic) metals geometric curvature leads to an orbital Rashba coupling and, as a direct byproduct, to a spin-Rashba coupling. 
Let us start by considering 
a single valley system with orbital degrees of freedom. For simplicity we will assume $p$ orbitals but the same arguments apply also to the $t_{2g}$ orbitals since they effectively span an effective angular momentum one subspace as the $p$ orbitals. 
The corresponding energy
spectrum is assumed to accurately represent the electronic bands close to the Fermi level of the
(magnetic) metal in question.
At the high-symmetry time-reversal invariant momentum around which the effective Hamiltonian is expanded and in the absence of geometric curvature, quantum confinement generally splits the three-dimensional irreducible representation (IRREP) corresponding to the atomic $p$ orbitals into a one-dimensional IRREP $a_{1g}$ and an $e_g$ doublet. To proceed further, let us obtain the effective Hamiltonian 
using the theory of invariants, which corresponds to a conventional ${\bf k} \cdot {\bf p}$ theory that keeps track of the point group symmetries of the crystal.
To do so, we use that a generic three-band Hamiltonian can be expanded in terms of the nine Gell-Mann matrices~\cite{mer23} $\Lambda_i$ as 
\begin{equation}
 {\mathcal H}({\bf k})=\sum_{i=0}^8 g_i({\bf k}) \Lambda_i  
\end{equation}
where $\Lambda_0$ corresponds to the identity. 
The  fact that the Hamiltonian must remain invariant requires that the momentum polynomials $g_i({\bf k})$ must belong to the same representation of the crystal point group of the corresponding Gell-Mann matrix. 
To make things concrete let us consider a crystalline systems with ${\mathcal D}_{4h}$ point group. The generators of the ${\mathcal D}_{4h}$ are a fourfold rotation ${\mathcal C}_{4v}$, a twofold rotation ${\mathcal C}_{2}^{\prime}$ with rotation axis orthogonal to the fourfold rotation axis, and the horizontal mirror $\sigma_h$. In Table~1 we list the representation of the Gell-Mann matrices in the $p_x,p_y,p_z$ orbital basis and momentum polynomials up to quadratic order in the ${\mathcal D}_{4h}$ point group. From this we find that the effective Hamiltonian away from the high-symmetry time-reversal invariant momentum reads 
%%%%%%%%%%%%%%%%%%%%
\begin{table}
\caption{Character tables\footnote{For simplicity we only report the generators of the group. Additionally, we indicate the representation of the Gell-Mann matrices and the polynomials of momentum ${\bf k}$ in two dimensions. The model Hamiltonians reported in the main text can be obtained by additionally using the time-reversal symmetry constraint ${\mathcal H}^{\star}(-k_x,-k_y)={\mathcal H}(k_x,k_y)$.}
%$^1$ 
for the point groups ${\mathcal D}_{4h}$ and ${\mathcal C}_{4v}$}
\begin{center}
\begin{tabular}{|c|c|c|c|c|c|c|c|}
\hline 
${\mathcal D}_{4h}$ & E & ${\mathcal C}_4$ & $ {\mathcal C}_2^{\prime}$ & $\sigma_h$  & $\mathcal{I}$ & polynomials of ${\bf k}$ & Gell-Mann matrices 
\tabularnewline
\hline 
$A_{1 g}$ & $1$ & $1$ & $1$ & $1$ & $1$ &  $1$,  $k_x^2 + k_y^2$ & $\Lambda_0$, $\Lambda_8$  
\tabularnewline 
\hline 
$A_{2 g}$ & $1$ & $1$ & $-1$ & $1$ & $1$ & --  & $\Lambda_2$ 
\tabularnewline  
\hline 
$B_{1g}$ & $1$ & $-1$ & $1$ & $1$ &  $1$ & $k_x^2- k_y^2$ & $\Lambda_3$
\tabularnewline  
\hline
$B_{2g}$ & $1$ & $-1$ & $-1$ & $1$ & $1$ & $k_x k_y$ & $\Lambda_1$ 
\tabularnewline  
\hline
$E_g$ & $2$ & $0$ & $0$ & $-2$ & $2$ & -- & $\left\{\Lambda_4, \Lambda_6 \right\} \hspace{.2cm} \left\{\Lambda_5, \Lambda_7 \right\} $ 
\tabularnewline 
\hline 
\end{tabular}

\vspace{1cm}
\begin{tabular}{|c|c|c|c|c|c|}
\hline 
${\mathcal C}_{4v}$ & E & ${\mathcal C}_{4}$ & $\sigma_v$ & polynomials of ${\bf k}$ & Gell-Mann matrices 
\tabularnewline
\hline 
$A_1$ & $1$ & $1$ & $1$ & $1$, $k_x^2+k_y^2$ & $\Lambda_0$, $\Lambda_8$  
\tabularnewline 
\hline 
$A_2$ & $1$ & $1$ & $-1$ & -- & $\Lambda_2$
\tabularnewline 
\hline
$B_1$ & $1$ & $-1$ & $1$ & $k_x^2-k_y^2$ & $\Lambda_3$
\tabularnewline 
\hline
$B_2$ & $1$ & $-1$ & $-1$ & $k_x k_y$ & $\Lambda_1$
 \tabularnewline 
\hline 
$E$ & $2$ & $0$ & $0$ & $\left\{k_x,k_y \right\} $ & $\left\{\Lambda_4, \Lambda_6 \right\} \hspace{.2cm} \left\{\Lambda_5, \Lambda_7 \right\} $ 
\tabularnewline  
\hline 
\end{tabular}
\end{center}
%\justifying
%\footnotesize{$^1$ For simplicity we only report the generators of the group. Additionally, we indicate the representation of the Gell-Mann matrices and the polynomials of momentum ${\bf k}$ in two dimensions. The model Hamiltonians reported in the main text can be obtained by additionally using the time-reversal symmetry constraint ${\mathcal H}^{\star}(-k_x,-k_y)={\mathcal H}(k_x,k_y)$} 
\end{table} 
%%%%%%%%%%%%%%%%%%%%%%%%
\begin{equation}
{\mathcal H}_{\bf k}=\Delta \, \Lambda_8 + \gamma_0 \Lambda_0 \left(k_x^2+k_y^2 \right) +\gamma_3 \left(k_x^2-k_y^2 \right) \Lambda_3 + \gamma_1 k_x k_y \Lambda_1, 
\end{equation}
where $\Delta$ indicates the crystal field splitting, $\gamma_0$ accounts for a conventional orbital independent while $\gamma_1$ and $\gamma_3$ parametrize $d$-wave orbital hybridization terms that yield orbital dependent effective masses. 
Let us now introduce geometric curvature in the thin film. It is well known~\cite{Landau_86} that across  the thin film curvature introduces a strain field that varies linearly across the film with regions under compressive and tensile strain separated by a mechanically neutral plane~\cite{Ortix2011,Stengel_2013}. 
The consequence of the presence of this strain field is twofold. 
First, the twofold rotation around the in-plane axis ${\mathcal C}_{2}^{\prime}$ is broken. Second, and perhaps even more importantly, the thin film will not possess any horizonal mirror symmetry. Therefore, even if the fourfold rotation symmetry is preserved, centrosymmetry is inevitably lost. 
%-- this absence of centrosymmetry is at the heart of the flexoelectric effect in insulators~\cite{Stengel_2013}. 
The highest symmetry point group symmetry will thus be the polar ${\mathcal C}_{4v}$ group. 
This geometry-induced symmetry reduction strongly changes the electronic structure of our multiorbital system. In fact, and as shown in Table~1, terms which are linear in momentum are allowed. The effective Hamiltonian up to linear order now reads
\begin{equation}
{\mathcal H}_{\bf k}=\Delta \, \Lambda_8 + \gamma_0 \Lambda_0 \left(k_x^2+k_y^2 \right) + \alpha_{OR} \left(k_x \Lambda_5 + k_y \Lambda_7 \right), 
\label{eq:hamorbitalRashba}
\end{equation}
where for simplicity we neglected the orbital hybridization terms  quadratic in momentum, and the linear term gets the form of the so-called orbital Rashba coupling~\cite{Park11,Kim13,Mer20} $k_x L_y - k_y L_x$ [see the Methods section]. Note that $k_{x,y}$ indicate now the momenta along the principal directions of the two-dimensional curved manifold.
Importantly, the curvature-induced orbital Rashba coupling is transmuted in an effective Rashba spin-orbit coupling (SOC) term in materials with a non-negligible atomic SOC [see the Methods section]. As a result, the effective Hamiltonian for a Kramers' doublet corresponds to the theory of a Rashba gas with an effective spin Hamiltonian 
\begin{equation}
{\mathcal H}_{\rm{spin}}({\bf k})  =
\alpha_R \,(k_x \sigma_y - k_y \sigma_x)+ \dfrac{\hbar^2 k^2}{2 m^{\star}} \sigma_0
\end{equation}
with the Rashba spin-orbit coupling strength $\alpha_R$ that is directly proportional to curvature-induced orbital Rashba coupling both when the atomic SOC energy is much larger than the crystal field splitting $\Delta$, and in the opposite weak spin-orbit coupling regime [see the Methods section]. 

Having obtained the form of the effective Hamiltonian in a multiorbital system when geometric curvature decreases the point group from ${\mathcal D}_{4h}$ to ${\mathcal C}_{4v}$, we next explicitly consider an effective model relevant for thin ferromagnetic shells. In doing so, we notice that the exchange coupling between the conducting electrons and the localized magnetic moments can be {\it a priori} influenced by curvature effects on the long-range magnetic order. It is well-known~\cite{Gaididei14,Kravchuk16}, in fact, that the magnetic energy functional in curved thin magnetic shell is endowed with a curvature-induced Dzyaloshinskii-Moriya interaction (DMI) and an effective magnetic anisotropy controlled by curvature. 
%The latter possesses two contribution different in nature. 
%Additionally, in the presence of an intrinsic DMI coupling results in a DMI-induced magnetic anisotropy. 
Generally speaking, these curvature-induced effects yield a reorganization of the magnetic texture. 
However, in a easy-surface magnetic nanomembrane such curvature-induced reorganization of the magnetic texture is absent~\cite{Ortix2023}. Consequently, the magnetic order parameter is  along the curved surface of the magnetic shell. 
The effective Hamiltonian for conduction electrons with the exchange coupling to localized moments then reads: 
\begin{equation}
{\mathcal H}^{\rm{spin}}_1({\bf k})  =
\alpha_R \,(k_x \sigma_y - k_y \sigma_x)+ \dfrac{\hbar^2 k^2}{2 m^{\star}} \sigma_0 + {\bf M} \cdot \boldsymbol{\sigma}
\label{eq:hamspinRashba}
\end{equation}
where ${\bf M}$ is the saturated magnetization and $\boldsymbol{\sigma}$ is the Pauli matrix vector. 

The Hamiltonian written above is universal and applies to any three-dimensional geometric shape the easy-surface thin magnetic shell acquires. It also implicitly use the fact that there are no strain fields along the principal in-plane curved directions. 
However, such inhomogeneous strain fields, experimentally shown in corrugated bilayer graphene~\cite{ho21}, are generally expected [see Fig.~\ref{fig:fig1}]. The presence of this strain fields can lead to a complete loss of rotational symmetries around the out-of-plane axis. 
%generally applicable to crystalline materials possessing an evenfold rotational symmetry. For thin films grown along a trigonal axis, a similar Hamiltonian can be obtained, but with an important difference. In the presence of the exchange coupling to the localized moments, {\it i.e.} when accounting for the broken time-reversal symmetry, the combined antiunitary symmetry ${\mathcal C}_2^{\prime} = {\mathcal C}_2 \Theta$ symmetry (with ${\mathcal C}_2$ the twofold rotation symmetry with rotation axis perpendicular to the magnetic shell and $\Theta$ the time-reversal symmetry operator) is broken. 
This implies that the spin Hamiltonian possesses an additional symmetry-allowed term $\lambda k_y \sigma_z$ where we assumed a residual mirror symmetry ${\mathcal M}_y$ sending $y \rightarrow -y$. 
The spin Hamiltonian in this case will read
\begin{equation}
{\mathcal H}^{\rm{spin}}_{2}({\bf k})  =
\alpha_R \,(k_x \sigma_y - k_y \sigma_x)+ \lambda k_y \sigma_z +\dfrac{\hbar^2 k^2}{2 m^{\star}} \sigma_0 + {\bf M} \cdot \boldsymbol{\sigma}
\label{eq:hamspinRashba3v}
\end{equation}

\section{Nonlinear anomalous Hall effect due to the quantum metric}

\begin{figure*}
    \centering
    \includegraphics[width=0.9\linewidth]{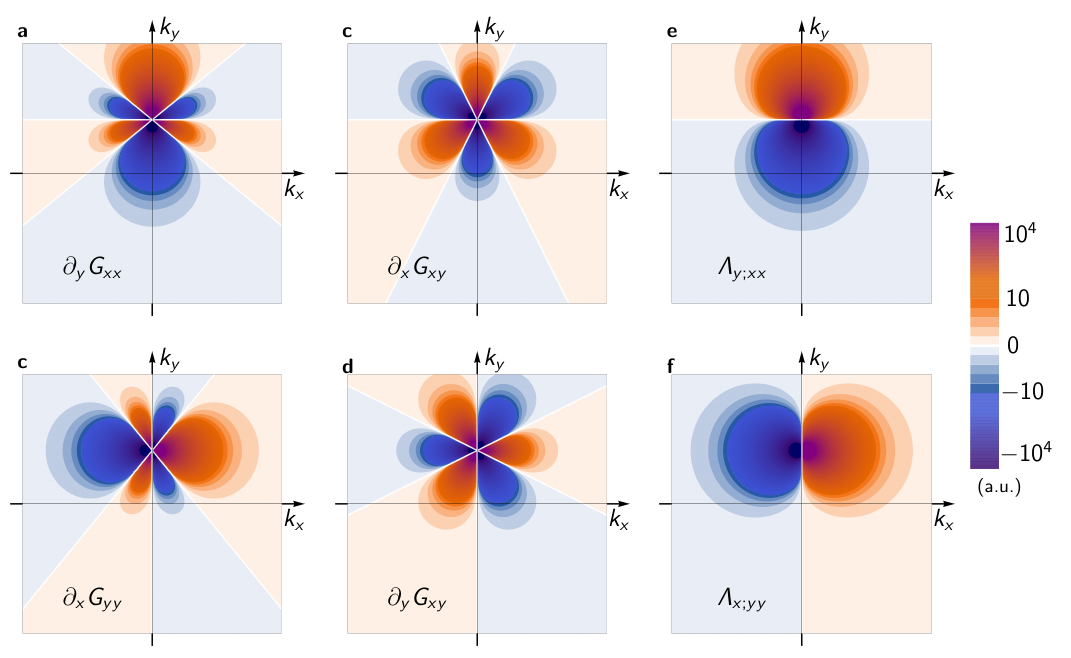}
\caption{{\bf Band Normalized Quantum Metric Dipoles.} {\bf a}-{\bf d} Contour plots of the band-normalized quantum metric (BNQM) dipoles $\partial_a G_{bc}$ (with $a,b,c,=x,y$ and $\partial_a\equiv\partial/\partial_{k_a}$) in momentum space in arbitrary units (a.u.) for a representative set of parameters and for  magnetization directed along $\hat x$. {\bf e}-{\bf f} Contour plots of the functions $\Lambda_{y;xx}(\mathbf{k})$ and $\Lambda_{x:yy}(\mathbf{k})$, respectively, which are built as linear combination of BNQM dipoles (see equation \eqref{eq:Lambda}). Notice that  $\Lambda_{y;xx}(\mathbf{k})$ is an even function of the momentum $k_x$ and will thus generally provide a non-vanishing contribution to the nonlinear anomalous Hall conductivity, while $\Lambda_{x;yy}(\mathbf{k})$ is an odd function of $k_x$ and the integration in k-space over the occupied state will vanish.
 \label{fig:fig2}}
\end{figure*}

The end product of the crystalline symmetry breaking due to bending is generally  the spin-momentum locking encoded in the Rashba spin-orbit coupling term of Eqs.~\eqref{eq:hamspinRashba},\eqref{eq:hamspinRashba3v}. It has been recently shown~\cite{Sala_24} that a prime physical consequence of spin-momentum locking is the fact that the corresponding Bloch waves are endowed with a non-trivial QGT. Importantly, the quantum metric is not forced to be vanishing at all momenta by the local in momentum ${\mathcal C}_2^{\prime}$ symmetry. This implies that both the Hamiltonians in Eqs.~\eqref{eq:hamspinRashba},\eqref{eq:hamspinRashba3v} have a non trivial QGT. Additionally, and as shown below, the symmetry properties of the quantum metric are the same in both cases.
%is allowed to be finite independent of the crystalline symmetries. In addition, the symmetries properties of the local concentrations of the quantum metric 
%This is in contrast with the properties of the imaginary part of the QGT correspo (see below).
%corresponding to the well-known Berry curvature. For 
%Importantly, while the real part of the QGT -- the so-called quantum metric -- is allowed to be finite, the imaginary part of the QGT that corresponds to the well-known Berry curvature is constrained to be vanishing. This is due to the presence of the antiunitary ${\mathcal C}_2^{\prime} = {\mathcal C}_2 \Theta$ symmetry with ${\mathcal C}_2$ the twofold rotation symmetry with rotation axis perpendicular to the membrane and $\Theta$ the time-reversal symmetry operator. Under ${\mathcal C}_2$ the out-of-plane component of the Berry curvature $\Omega_z(k_x,k_y) \rightarrow \Omega_z(-k_x,-k_y)$. Additionally, time-reversal symmetry sends $k_{x,y} \rightarrow -k_{x,y}$ and concomitantly reverse the Berry cruvature $\Omega_z \rightarrow -\Omega_z$. Therefore we obtain that the combined $\mathcal{C}_2^{\prime}$ symmetry implies $\Omega_z(k_x,k_y)= -\Omega_z(k_x,k_y) =0$. We note that ${\mathcal C}_2^{\prime}$ is inherently present in a easy-surface magnetic shell thereby guaranteeing that the Berry curvature and its related transport effects cannot appear even if time-reversal symmetry is broken.

%Let us now consider the symmetry constraints on the non-vanishing quantum metric 
The quantum metric is defined by 
$g_{ab}(k_x,k_y)= \partial_{k_a} \hat{\bf d} \cdot \partial_{k_b} \hat{{\bf d}} / 4 $ where $\hat{{\bf d}}={\bf d} / |{\bf d}|$ is the normalized Hamiltonian vector defined by ${\mathcal H}^{\rm spin}({\bf k})= \bf{d} \cdot \boldsymbol{\sigma}  $. Without loss of generality, we fix the direction of the saturated magnetization along the $\hat{x}$ axis of the thin magnetic shell: recall that in the presence of bending $\hat{x}$ corresponds to one of the principal direction of the curved membranes shape. 
For systems with an evenfold rotation symmetry in the non-magnetic phase, and thus described by Eq.~\eqref{eq:hamspinRashba}, the magnetic point group contains the combined ${\mathcal M}_y^{\prime}={\mathcal M}_y \Theta$ symmetry with ${\mathcal M}_y$ the vertical mirror symmetry parallel to the magnetization, 
and a residual ${\mathcal M}_x$ symmetry. 
On the contrary systems described by Eq.~\eqref{eq:hamspinRashba3v} have a magnetic point group containing only ${\mathcal M}_y$. 
%We note that when considering the saturated magnetization along the $\hat{y}$ direction, the magnetic point group of Eq.~\eqref{eq:hamspinRashba3v} has only a residual ${\mathcal M}_y$ symmetry. 
Under both ${\mathcal M}_y^{\prime}$ and ${\mathcal M}_x$ we have that the crystalline momentum $k_x \rightarrow -k_x$ whereas $k_y \rightarrow k_y$. By additionally using the transformation of the symmetric quantum metric tensor under the vertical mirror symmetry we then get that the following symmetry constraints
\begin{eqnarray}
g_{x x}(k_x,k_y)&=& g_{x x}(-k_x,k_y) \nonumber \\
g_{y y}(k_x,k_y)&=& g_{y y}(-k_x,k_y)  \\
g_{x y}(k_x,k_y)&=& -g_{x y}(-k_x,k_y). \nonumber
\end{eqnarray}
Therefore, we can conclude that the properties of the quantum metric are independent of the specific model Hamiltonian choice.

We next show how these symmetry-enforced properties of the quantum metric are reflected in the nonlinear electron transport characteristics, specifically on the electric nonlinear anomalous Hall effect induced by the quantum metric. The latter is in fact governed by the dipoles of the band-energy normalzed quantum metric (BNQM) $G^{\pm}_{a b}(k_x,k_y)=\pm g_{a b}(k_x,k_y) / |{\bf d}|$, where the $\pm$ refers to the internal and external Rashba electronic branches respectively. Specifically we have that the intrinsic transversal nonlinear conductivity components~\cite{gao14,das23,kap24} read
\begin{equation}
\sigma_{a;bb}^{QM}=-\dfrac{e^3}{\hbar}\sum_n \int \dfrac{d^2{\bf k}}{(2 \pi)^2} \,\Lambda_{a;bb}^{n}({\bf k}) f_n({\bf k}) 
\end{equation}
where $n=\pm$, $f_n$ indicates the Fermi-Dirac distribution function, and the BNQM dipole densities
\begin{equation}
\label{eq:Lambda}
\Lambda_{a;bb}^{\pm}({\bf k})=\dfrac{3}{2} \partial_a G^{\pm}_{bb}({\bf k}) -\partial_b G_{a b}^{\pm}({\bf k}). 
\end{equation}

In Fig.~\ref{fig:fig2} we show the momentum space maps of the two BNQM dipole densities governing the nonlinear anomalous Hall conductivities $\sigma_{y;xx}$ and $\sigma_{x;yy}$
assuming the model Hamiltonian in Eq.~\eqref{eq:hamspinRashba} -- we refer to the Supplementary Information for an analogous discussion considering Eq.~\eqref{eq:hamspinRashba3v}.
The BNQM dipole density $\Lambda_{y;xx}({\bf k})$ is enforced by symmetry to be an even function of the momentum $k_x$ and will thus generally provide a non-vanishing contribution to the nonlinear anomalous Hall conductivity $\sigma_{y;xx}$. The BNQM dipole $\Lambda_{x;yy}({\bf k})$ instead is forced by symmetry to be an odd function of the momentum $k_x$. Together with the fact that the electronic band energies $\epsilon_{\pm}(k_x,k_y)=\epsilon_{\pm}(-k_x,k_y)$ by virtue of ${\mathcal M}_y^{\prime}$ and/or ${\mathcal M}_x$ we thus conclude that the non-trivial quantum metric does not a generate a nonlinear anomalous Hall conductivity when the driving electric field is orthogonal to the in-plane magnetization. Similar symmetry constraints are also found when considering the extrinsic nonlinear Drude contribution that generally reads 
%%%%%%%%%%%
\begin{equation}
     \sigma_{a;bb}^{NLD} = %frac{e^3\tau^2}{\hbar^3}\sum_{n\in occ.}\int \frac{d^2 \mathbf{k}}{(2\pi)^2} \partial_a\varepsilon_n(\mathbf{k}) \partial_b\partial_c f_n(\mathbf{k})=
     -\frac{e^3\tau^2}{\hbar^3}\sum_{n\in occ.}\int \frac{d^2 \mathbf{k}}{(2\pi)^2} \partial_a\partial_b\partial_b\varepsilon_n(\mathbf{k})f_n(\mathbf{k}),
\end{equation} 
%%%%%%%%
where $\tau$ indicates the electronic scattering time. 
The band energy derivative $\partial_{k_x} \partial_{k_y} \partial_{k_y} \epsilon_{\pm}({\bf k})$ is enforced to be an odd function of $k_x$ thus guaranteeing  the absence of the nonlinear Drude contribution $\sigma_{x; yy}^{NLD}$. On the contrary, the nonlinear Drude contribution $\sigma_{y; xx}^{NLD}$ is not forced to vanish by symmetry and will thus coexist with the intrinsic quantum metric contribution discussed before.

\begin{figure*}
    \centering
    \includegraphics[width=0.7\linewidth]{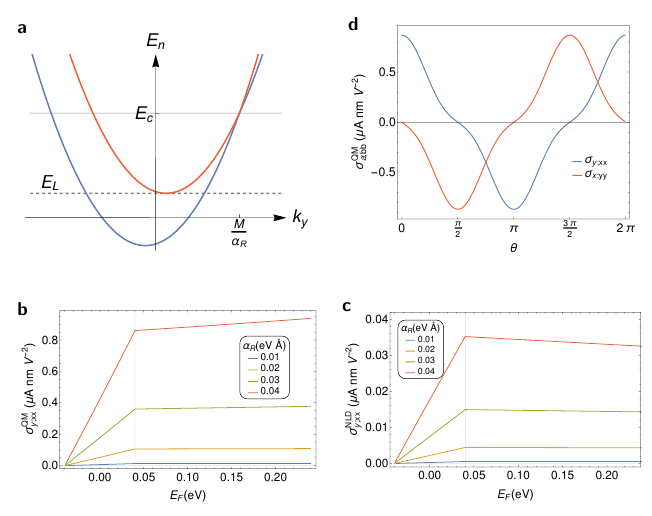}
\caption{{\bf Quantum metric induced nonlinear anomalous Hall effect.}  {\bf a} Energy bands as a function of $k_y$ for $k_x=0$ when the magnetization $\mathbf{M}$ is in the $\hat{x}$ direction. The dashed line marks the energy of the Lifshitz transition $E_L$. The band-crossing occurs at $(k_x=0;k_y^c=M/\alpha_R)$  and $E_c=\hbar^2 M^2/ 2 m^\star \alpha_R^2$.  {\bf b} Behavior of the quantum metric induced nonlinear conductivity tensor component $\sigma^{QM}_{y;xx}$  as a function of the Fermi energy $E_F$ for $\mathbf{M}=40\,\hat{x}$~meV and four different values of Rashba spin-orbit coupling $\alpha_R$ (see legend). The kink in the behavior occurs for $E_F=E_L$. {\bf c} Plot of the nonlinear Drude conductivity tensor component $\sigma^{NLD}_{y;xx}$ for the same set of parameters of {\bf b} and relaxation time $\tau=10$fs. {\bf d} Plot of %quantum metric induced transverse conductivity
$\sigma_{a;bb}^{QM}$  (with $\{a,b\}=\{x,y\}$) as a function of the in plane orientation $\theta$ of the magnetization for $M=40$meV, $\alpha_R=0.04$eV\r{A} and $E_F=80$meV. The two components $(y;xx)$ and $(x;yy)$ are related by a $\pi/2$ shift and show a $2\pi$-periodic non-sinusoidal behavior.
    \label{fig:fig3}}
\end{figure*}

Next, we discuss the behavior of the nonlinear Drude and the quantum metric-induced contributions to the nonlinear conductivity $\sigma_{y; xx}$. First, we note that the energy spectrum of the model Hamiltonian in Eq.~\eqref{eq:hamspinRashba} features a crossing protected by ${\mathcal M}_x$ on the mirror symmetric line $k_x=0$ [see Fig.~\ref{fig:fig3}(a)]. The finite $k_y$ value at which the crossing occurs is determined by $k_y^c=M/\alpha_R$ and shifts away from the time-reversal invariant momentum as the magnetization increases. The corresponding crossing energy $E_c=\hbar^2 M^2 / (2 m^{\star} \alpha_R^2)$ identifies two different regimes for the properties of the Fermi lines [see the Supplementary Information]. In fact, at Fermi energies $E_F > E_c$ the two Fermi lines corresponding to the two spin bands wind around the crossing point ${\bf k}^c = \left\{0,k_y^c\right\}$.
%\textcolor{blue}
This winding does not occur for $E_F \leq E_c$. Additionally, in this regime
%When $E_F \leq E_c$
%On the contrary, for $E_F<E_c $ the two Fermi lines wind around the time-reversal invariant momentum. 
%Importantly, in this regime 
we find the occurrence of a Lifshitz transition determined by $E_F \equiv E_L$ [see Fig.~\ref{fig:fig3}(a)] below which a single spin band is  occupied [see also the Supplementary  Information].

In Fig.~\ref{fig:fig3}(b) we show the Fermi energy dependence of the intrinsic quantum metric induced nonlinear anomalous Hall conductivity term $\sigma^{QM}_{y;xx}$. For $E_F<E_L$ we find that $\sigma^{QM}_{y;xx}$ grows linearly with a slope that increases with the Rashba coupling $\alpha_R^3$. At the Lifshitz transition, the nonlinear conductivity displays a kink that is followed by a much weaker and nonlinear Fermi energy dependence. By further increasing $E_F$ (see the Supplementary Information) one finds that the zero-temperature nonlinear conductivity is characterized by an asymptotic behavior at $E_c$ that resembles the behavior of the nonlinear magnetoconductivity $\sigma_{x;xx}$
in nonmagnetic two-dimensional electron gases~\cite{Sala_24}. The extrinsic nonlinear Drude contribution $\sigma_{y;xx}^{NLD}$ is characterized by features [see Fig.~\ref{fig:fig3}(c)] which are very similar to $\sigma_{y;xx}^{QM}$ only in the $E_F<E_c$ regime. In fact, in the opposite $E_F>E_c$ regime the nonlinear Drude contribution vanishes thus making the anomalous Hall conductivity $\sigma_{y;xx}$ entirely intrinsic. 
Additionally, in the $E_F<E_c$ regime the two contributions can be parsed thanks to the different dependence on the electronic scattering time $\tau$. The latter can be controlled using the temperature dependence of the linear conductivity. Transport experiments at variable temperature will thus allow to single out the quantum metric-induced contribution to the nonlinear conductivity.

In Fig.~\ref{fig:fig3}(d) we show the angular dependence of the intrinsic conductivity $\sigma_{y;xx}^{QM}$ as obtained by tilting the magnetic field direction from $\hat{x}$ by a $\theta$ angle.  We find that the nonlinear conductivity is $2 \pi$-periodic in agreement with the fact that $\sigma_{y;xx}^{QM}$ must be odd in the magnetization $M$. Additionally, due to the breaking of the ${\mathcal M}_x,{\mathcal M}_y^{\prime}$ symmetries the nonlinear conductivity $\sigma_{x;yy}$ acquires finite values at $\theta \neq 0$. However, the angular dependence of the two nonlinear conductivities are related to each other by a simple $\pi/2$ offset. This signals that, independent of the magnetization direction, when the driving electric field is collinear to the magnetization, only a transversal nonlinear current will be generated. We emphasize that this property is specific to our model Hamiltonian in Eq.~\eqref{eq:hamspinRashba}, which, does not feature crystalline anisotropy terms and thus possess full rotational symmetry in the absence of magnetization [see also Supplementary Information].

\section{Berry curvature-induced nonlinear anomalous Hall effect}
In this section, we will show that thin magnetic shells where geometric deformations lead to a complete loss of rotational symmetries, there is an additional contribution to the nonlinear Hall conductivity induced by the Berry curvature. 
We start by recalling that 
under a ${\mathcal C}_2$ rotation the out-of-plane component of the Berry curvature $\Omega_z(k_x,k_y) \rightarrow \Omega_z(-k_x,-k_y)$. Additionally, time-reversal symmetry sends $k_{x,y} \rightarrow -k_{x,y}$ and concomitantly reverse the Berry curvature $\Omega_z \rightarrow -\Omega_z$. Therefore we obtain that the combined $\mathcal{C}_2^{\prime}$ symmetry implies $\Omega_z(k_x,k_y)= -\Omega_z(k_x,k_y) =0$. We note that ${\mathcal C}_2^{\prime}$ is inherently present in systems described by the Hamiltonian in Eq.~\eqref{eq:hamspinRashba} thereby guaranteeing that the Berry curvature and its related transport effects cannot appear even if time-reversal symmetry is broken.

\begin{figure*}
    \centering
    \includegraphics[width=0.95\linewidth]{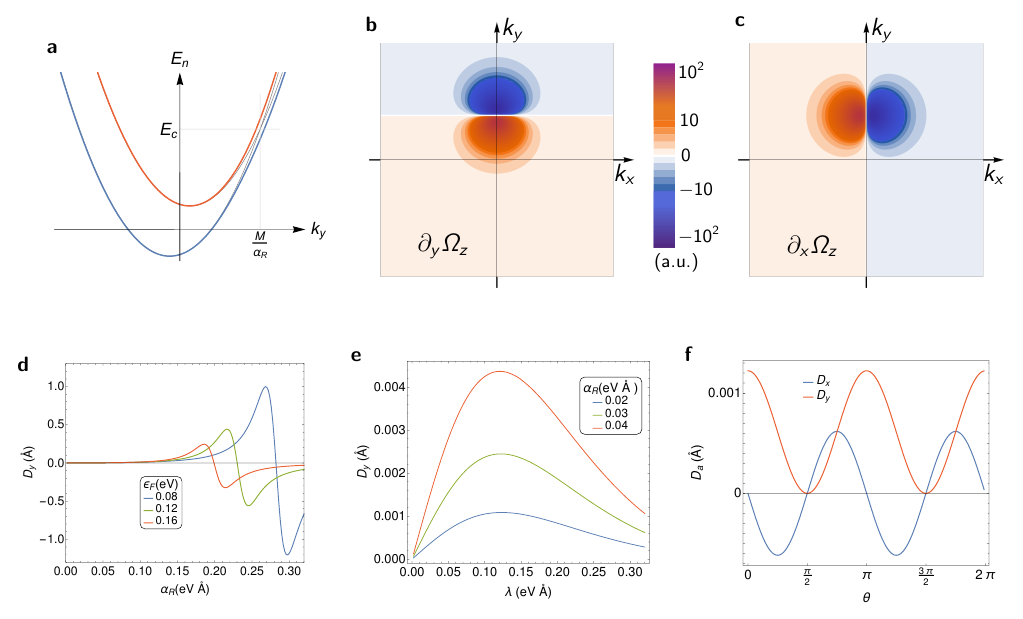}
\caption{{\bf Berry curvature induced nonlinear anomalous Hall effect.}  
{\bf a} Energy bands as a function of $k_y$ for $k_x=0$ when the magnetization is in the $\hat{x}$ direction and for $\lambda \neq 0$. The coupling $\lambda$ removes the degeneracy and, as reference, the $(\lambda=0)$-energy bands are plotted in gray. {\bf b}-{\bf c} Contour plots of the k-space gradient of the Berry curvature $\partial_y\Omega_z$ and $\partial_y\Omega_z$, respectively,
%with $\partial_a\equiv\partial/\partial k_a$, 
for a representative set of parameters and  $\mathbf{M}\parallel \hat{x}$. 
{\bf d-e} Behavior of the Berry curvature dipole (BCD)  $D_y$  as a function of the Rashba coupling $\alpha_R$ and of parameter $\lambda$, respectively, for $\mathbf{M}=40\,\hat{x}$meV. %with $\theta=0$.
In {\bf d} the BCD is displayed for $\lambda=0.02$eV\r{A} and for three different values of the Fermi energy. The BCD features a sign change and the double peak structure which occur for values of $\alpha_R$ close to the condition $E_F=E_c$. In
{\bf e} the Fermi energy is set to $E_F=80$meV and the BCD shows a non-monotonic behavior as a function of $\lambda$. In 
{\bf f} the BCDs $D_x$ and $D_y$ are presented as a function of the angle $\theta$ formed by the magnetization $\mathbf{M}$ and the $\hat{x}$ axis. The BCDs show a $\pi$-periodic behavior, in contrast with the quantum metric induced nonlinear conductivity. Notice that with the choice $\lambda k_y \sigma_z$ the Berry curvature is zero for $\theta=\pi/2$. The parameters used here are $M=40$meV, $\alpha_R=0.04$eV\r{A} $E_F=80$meV, $\lambda=0.02$eV\r{A}.
    \label{fig:fig4}}
\end{figure*}

%In this section, we will show that in systems without rotational symmetries an additional contribution to the nonlinear Hall conductivity induced by the Berry curvature exists. The Berry curvature-mediated nonlinear transport is maximal when the driving electric field is orthogonal to the magnetization and is thus fundamentally different from the quantum metric-induced contribution that is instead forced to vanish in this configuration. 

%\textcolor{red}{Esplicitare meglio i casi and riferirsi solo a Myprime}

%We start by noticing that the absence of rotational symmetry implies that either the antimirror ${\mathcal M}_{y}^{\prime}$ or the vertical ${\mathcal M}_x$ mirror symmetry is broken. The latter situation can be captured by adding a mirrro-symmetry allowed $\lambda k_x \sigma_z$ to the spin Hamiltonian in Eq.~\eqref{eq:hamspinRashba}. However, the Berry curvature defined by $\Omega_z^{\pm}= \mp \frac12 \hat{\mathbf{d}} \cdot ( \partial_{k_x} \hat{\mathbf{d}} \times \partial_{k_y} \hat{\mathbf{d}})$ is still vanishing even though the local ${\mathcal C}_2 \Theta$ symmetry is explicitly broken. 

%Note that this is in agreement with the fact that the vertical mirror symmetry still preserved a crossing on the $k_x=0$ mirror symmetric line. Therefore, in this situation only a quantum metric-induced nonlinear anomalous Hall conductivity $\sigma_{y;xx}^{QM}$ can appear. 

For systems with a residual ${\mathcal M}_y^{\prime}$ symmetry, and thus described by the Hamiltonian in Eq.~\eqref{eq:hamspinRashba3v} with the saturated magnetization that is fixed again along the $\hat{x}$ direction, the situation is completely different. 
%The loss of rotations can be captured by adding a $\lambda k_y \sigma_z$ term to the Hamiltonian in Eq.~\eqref{eq:hamspinRashba}. 
The effect of the $\lambda k_y \sigma_z$ term is twofold. First, it removes the mirror-symmetry protected crossing with the two spin bands that become non-degenerate at all ${\bf k}$ points [see Fig.~\ref{fig:fig4}(a)]. Second, it endows the system with a local Berry curvature with non-zero average. 
This also implies that, 
%in perfect anology with the ferromagnetic oxide  SrRuO$_3$ and the Kagome magnet Fe$_3$Sn$_2$, 
the system features an in-plane anomalous Hall effect, which represents the magnetic analog of the anomalous planar Hall effect predicted in two-dimensional systems~\cite{Battilomo2021,Cullen2021}.
%and experimentally observed at LaAlO$_3$/SrTiO$_3$ heterointerfaces, in heterodimsensional superlattices, and, more recently in the Weyl semimetal PtBi$_2$. 
Perhaps even more importantly, the momentum dependence of the Berry curvature is characterized by a non-vanishing Berry curvature dipole~\cite{Sodemann2015,Ortix2021,Du21}
\begin{equation}
   D_y=\sum_n \int \dfrac{d^2{\bf k}}{(2 \pi)^2} \partial_{k_y} \Omega_z^n({\bf k}) f_n({\bf k}).
\end{equation}
The residual ${\mathcal M}_y^{\prime}$ symmetry implies in fact that the Berry curvature $\Omega_z^n(k_x,k_y)=\Omega_z^n(-k_x,k_y)$. 
The dipole density $\partial_{k_y} \Omega_z^n({\bf k})$ is consequently an even function of $k_x$ [see Fig.~\ref{fig:fig4}(b)], as the band energies are, thereby implying a non-vanishing dipole $D_y$. On the contrary, the dipole $D_x$ is forced to vanish for the simple reason that its density [see Fig.~\ref{fig:fig4}(c)] is an odd function of $k_x$. 

The behavior of the BCD $D_{y}$ as a function of the Rashba spin orbit coupling constant $\alpha_R$ is displayed in Fig.~\ref{fig:fig4}(d) for different values of the Fermi energy. It features a double-peak structure occurring  
when the Fermi energy sits in vicinity of the avoided level crossing at $E_c$. This is because in this region of near degeneracy the Bloch waves are rapidly changing, and thus produce hotspots of BC and BCD. Fig.~\ref{fig:fig4}(e) shows the behavior of $D_y$ by increasing the strength of the mirror symmetry breaking term parametrized by $\lambda$. The dipole has a non-monotonic behavior and is maximized at a finite $\lambda$ value that is weakly dependent on the Rashba spin-orbit coupling strength. The fact that the dipole displays this non-monotonic dependence can be understood by noticing that in the $\lambda \rightarrow 0$ limit, the dipole must vanish since the ${\mathcal C}_2 \Theta$ symmetry is restored. On the other hand, the dipole vanishes also in the $\lambda \rightarrow \infty$ limit. This can be shown using the following argument: taking only the leading order terms, the Hamiltonian in Eq.~\eqref{eq:hamspinRashba3v} can be written as
%with the addition of the mirror symmetry breaking term $\lambda k_y \sigma_z$ can be written as 
\begin{equation}
    {\mathcal H}^{\rm{spin}}_2({\bf k}) \simeq \alpha_R k_x \sigma_y + M \sigma_x  + \lambda  k_y \sigma_z + \dfrac{\hbar^2 k^2}{2 m^{\star}} \sigma_0
    \label{eq:hamapprox}
\end{equation}
The first three terms in the equation above, which determine the Berry curvature, describe an anisotropic massive Dirac cone with Fermi velocities $\alpha_R$ and $\lambda$ and mass given by the magnetization $M$. The corresponding Berry curvature becomes strongly peaked at $k_y=0$ as the related velocity $\lambda$ increases, which justifies the use of Eq.~\eqref{eq:hamapprox} in the $\lambda \rightarrow \infty$ limit. Furthermore, both the approximate Berry curvature and the band energies become even functions of the momentum $k_y$ thereby implying that the corresponding dipole $D_y$ must vanish. 

The existence of a finite Berry curvature dipole $D_y$ at finite values is reflected in the appearance of a non-vanishing nonlinear conductivity $\sigma_{x;yy}=e^3 \tau / \hbar^2 D_y$. Note that the linear in $\tau$ dependence is required in order to have a $\sigma_{x;yy}$ compatible with the ${\mathcal M}_y^\prime$ symmetry. The linear dependence in the relaxation time also implies that the magnetization dependence of $\sigma_{x;yy}$ must be an even function, contrary to $\sigma_{y;xx}$ that is odd in ${\bf M}$. As a result, a swap of the magnetization does not change the Berry curvature dipole-induced nonlinear conductivity. This is immediately demonstrated in Fig.~\ref{fig:fig4}(f) where we show the behavior of the dipole $D_{y}$ by changing the magnetization angle with respect to the $\hat{x}$ axis. We have that the Berry curvature dipole has a $\pi$ periodicity. Note that when the magnetization direction $\theta \neq 0$ the Berry curvature dipole acquires a $\hat{x}$ component. Both dipoles however disappear at $\theta \equiv \pi/2$. This is due to  the fact that in this configuration the system preserves the vertical mirror symmetry ${\mathcal M}_y$ and a crossing on the $k_y=0$ mirror symmetric line will therefore occur. The Berry curvature itself in this situation will therefore vanish.

\section{Conclusions}
To wrap up, we have shown that easy-surface thin ferromagnetic shells even of centrosymmetric materials will display a nonlinear anomalous Hall effect when acquiring curved geometric shapes. Despite the absence of curvature effects on the magnetic order parameter, curvature drives non-trivial quantum geometric properties of the Bloch waves via the occurrence of spin-momentum locked textures. We have shown that when undergoing bending, easy-surface ferromagnetic shells will then display a nonlinear anomalous Hall effect that is maximized  when the driving electric field is collinear to the saturated magnetization. Additionally, whenever geometric deformations lead to a complete rotational symmetry breaking a complementary nonlinear anomalous Hall effect due to the Berry curvature dipole can be observed when the driving electric field is orthogonal to the magnetization. The combination of these two effects therefore allow to obtain a complete mapping of the quantum geometric tensor using nonlinear transport experiments. 
Our theoretical study thus shows real-space geometric control of the quantum geometric properties of the electronic wavefunctions in magnetic materials and highlights a novel spin-orbitronic effect in magnetic nanomembranes. 

%the mirror symmetry breaking term $\lambda k_y \sigma_z$ in the Hamiltonian of Eq.~\eqref{eq:hamspinRashba} plays the role of a $k_y$ dependent Dirac mass that vanishes on the $k_y=0$ line. In the $\lambda \rightarrow \infty$ limit, the Berry curvature will be strongly localized on the $k_y=0$ line.
%thus be strongly peaked close to the $k_y=0$ line and 
%Let us consider the Hamiltonian in Eq.~\eqref{eq:hamspinRashba} including the mirror symmetry breaking term $\lambda k_y \sigma_z$ and introduce the shifted momenta $p_y=k_y-M/\alpha_R$ and $p_x=k_x$. The Hamiltonian can be then written as 
%\begin{equation}
 %   {\mathcal H}_{\rm{spin}}({\bf k})=\alpha_R(p_x \sigma_y - p_y \sigma_x) + \lambda \sigma_z \left(\dfrac{M}{\alpha_R} + p_y\right) + \dfrac{\hbar^2}{2 m^{\star}} \left[p_x^2 + %(p_y+\dfrac{M}{\alpha_R})^2 \right] \sigma_0
%\end{equation}
%At the 

\section{Methods}

\subsection{Representation of the Gell-Mann matrices}
The  eight Gell-Mann matrices can be defined as 
\begin{eqnarray*}
\Lambda_{1}=\begin{pmatrix}0 & 1 & 0\\
1 & 0 & 0\\
0 & 0 & 0
\end{pmatrix}, & \Lambda_{2}=\begin{pmatrix}0 & -i & 0\\
i & 0 & 0\\
0 & 0 & 0
\end{pmatrix},\\
\Lambda_{3}=\begin{pmatrix}1 & 0 & 0\\
0 & -1 & 0\\
0 & 0 & 0
\end{pmatrix}, & \Lambda_{4}=\begin{pmatrix}0 & 0 & 1\\
0 & 0 & 0\\
1 & 0 & 0
\end{pmatrix},\nonumber \\
\Lambda_{5}=\begin{pmatrix}0 & 0 & -i\\
0 & 0 & 0\\
i & 0 & 0
\end{pmatrix}, & \Lambda_{6}=\begin{pmatrix}0 & 0 & 0\\
0 & 0 & 1\\
0 & 1 & 0
\end{pmatrix},\nonumber \\
\Lambda_{7}=\begin{pmatrix}0 & 0 & 0\\
0 & 0 & -i\\
0 & i & 0
\end{pmatrix}, & \quad\Lambda_{8}=\begin{pmatrix}\tfrac{1}{\sqrt{3}} & 0 & 0\\
0 & \tfrac{1}{\sqrt{3}} & 0\\
0 & 0 & \tfrac{-2}{\sqrt{3}}
\end{pmatrix}.\nonumber 
\end{eqnarray*}
%and $\Lambda_0$ is the identity matrix.

\subsection{Derivation of the Rashba Hamiltonian}
To derive an effective spin Hamiltonian, we first note that the conventional angular momentum operators are related to the Gell-Mann matrices introduced above by $\Lambda_2=L_z$, $\Lambda_7=L_x$, and $\Lambda_5=-L_y$. This implies that the term linear in momentum in the Hamiltonian of Eq.~\eqref{eq:hamorbitalRashba} can be written as 
\begin{equation}
H_{\rm OR}
=
\alpha_{OR}( k_x L_y - k_y L_x ).
\end{equation}
Let us now include the spin-orbit coupling term in the Hamiltonian
$
H_{\rm SO}
=
\lambda_{SOC}\,\mathbf{L}\cdot\mathbf{S}
=
 \lambda_{SOC}\sum_{\alpha=x,y,z} L_\alpha \sigma_\alpha/2$. 
In the weak crystal field splitting regime $\Delta \ll \lambda_{SOC}$, we have that that the spin-orbit coupling splits the (effective) $p$-orbital manifold into a $J=3/2$ quartet at energy $+\lambda_{SOC}/2$ and a $J=1/2$ doublet at energy $-\lambda_{SOC}$. The total energy splitting is therefore $\Delta = E_{3/2} - E_{1/2} = \frac{3}{2}\lambda_{SOC}.$
We next derive the effective low-energy model for the $J=1/2$ doublet. To this aim, we recal that that $\ket{J,J_z}$ states can be expressesed in terms of the $\ket{L,L_z} \otimes \ket{\sigma,\sigma_z}$ states using the Clebsch-Gordan coefficients as 
%Let us consider the Hamiltonian
%\begin{equation}
%H = H_{\rm cOR} + H_{\rm SO},
%\end{equation}
%with bent induced orbital Rashba coupling
%\begin{equation}
%H_{\rm cOR}
%=
%\alpha_{xy}( k_x L_y - k_y L_x )
%+ \alpha_{z} (k_x + k_y) L_z,
%\end{equation}
%and atomic spin--orbit coupling 
%\begin{equation}
%H_{\rm SO}
%=
%\lambda_{SOC}\,\mathbf{L}\cdot\mathbf{S}
%=
%\frac{\lambda_{SOC}}{2}\sum_{\alpha=x,y,z} L_\alpha \sigma_\alpha.
%\end{equation}
%%
%In the presence of the spin-orbit coupling the $p$-manifold splits into:
%\begin{itemize}
%\item a $J=3/2$ quartet at energy $+\lambda_{SOC}/2$,
%\item a $J=1/2$ doublet at energy $-\lambda_{SOC}$.
%\end{itemize}
%%
%The energy splitting is
%\begin{equation}
%\Delta = E_{3/2} - E_{1/2} = \frac{3}{2}\lambda_{SOC}.
%\end{equation}
%In order to construct the effective low energy model for the J=1/2 doublet we project the orbital Rashba Hamiltonian in that subspace.
%To this aim we recall that the total angular momentum
%eigenstates for $J=\tfrac12$ are
\begin{align*}
\left| \tfrac12,\,+\tfrac12 \right\rangle
&=
\sqrt{\frac{2}{3}}\,
|1,1\rangle\,\left|\tfrac12,-\tfrac12\right\rangle
-
\sqrt{\frac{1}{3}}\,
|1,0\rangle\,\left|\tfrac12,\tfrac12\right\rangle,
\\[6pt]
\left| \tfrac12,\,-\tfrac12 \right\rangle
&=
\sqrt{\frac{2}{3}}\,
|1,-1\rangle\,\left|\tfrac12,\tfrac12\right\rangle
-
\sqrt{\frac{1}{3}}\,
|1,0\rangle\,\left|\tfrac12,-\tfrac12\right\rangle.
\end{align*}
with the angular momentum eigenstates that are related to the atomic real orbitals by 
$\ket{1,\pm 1}=\mp (p_x \pm i p_y) / \sqrt{2}$ and $\ket{1,0}=p_z$. The two states composing the $J=1/2$ doublet 
$\ket{1/2, \pm 1/2}$ are related to the real orbitals by 
%\[
%|1,1\rangle = -\frac{1}{\sqrt{2}}(p_x %+ i p_y), 
%\qquad
%|1,0\rangle = p_z,
%\qquad
%|1,-1\rangle = \frac{1}{\sqrt{2}}(p_x - i p_y) \,.
%hy7\]
%Taking into account the expressions of the $p$-orbitals, one can express the $J=1/2$ doublet configurations as 
\[
{
\left| \tfrac12,\,+\tfrac12 \right\rangle
=
-\frac{1}{\sqrt{3}}
\left(
p_z\,|\uparrow\rangle
+ p_x\,|\downarrow\rangle
+ i\,p_y\,|\downarrow\rangle
\right)
}
\]
\[
{
\left| \tfrac12,\,-\tfrac12 \right\rangle
=
-\frac{1}{\sqrt{3}}
\left(
p_z\,|\downarrow\rangle
- p_x\,|\uparrow\rangle
+ i\,p_y\,|\uparrow\rangle
\right)
}
\]
Let us now introduce the effective spin-$1/2$ operator $\tau$ spanning the $J=1/2$ space. 
%Hence, one can introduce a pseudospin $\tau$ acting in subspace$\{\left|+\right\rangle, \left|-\right\rangle\}$, where $\left|+\right\rangle=\left| \tfrac12,\,+\tfrac12 \right\rangle$ and $\left|-\right\rangle=\left| \tfrac12,\,-\tfrac12 \right\rangle \}$.
By evaluating the angular momentum operators over the states introduces above we have that in the $J=1/2$ basis 
$L_i \rightarrow \frac{1}{3} \tau_i$. 
Note that the projection of terms of the form $L_i L_j$ (which include the orbital hybridization terms quadratic in momentum) %\textcolor{red}{please check $L_i L_j=\Lambda_{3,1}$ of the character tables} 
can be expressed as 
%One can immediately show that projected $L_i$ can be expressed as $\Tilde{L}_i=\frac{1}{3} \tau_i$.
%For completeness, one can also show that the projected quadrupole terms $L_i L_j$ in the projected subspace can be expressed as 
$L_i L_j \rightarrow \frac{1}{9} \tau_i \tau_j$ which however yields only an identity term in the effective Hamiltonian as a result of the relation for the Pauli matrices 
%only identity contributions in the Hamiltonian due to the following algebra relation of the Pauli matrices 
$\tau_\alpha \tau_\beta
=
\delta_{\alpha\beta}
+
i\,\epsilon_{\alpha\beta\gamma}\,\tau_\gamma$ with $\delta_{\alpha\beta}$ the Kronecker delta. As a result, we have that in the strong spin-orbit coupling regime the effective Hamiltonian reads
\begin{equation}
\Tilde{\mathcal H}_{}
=
\frac{1}{3} \alpha_{OR}( k_x \tau_y - k_y \tau_x )
\end{equation}
with the characteristic Rashba spin-orbit coupling strength that is a third of the orbital Rashba coupling.

That the concomitant presence of orbital Rashba coupling and spin-orbit coupling gives rise to an effective Rashba spin-orbit coupling can be also seen in the opposite limit in which the atomic spin-orbit couplins is considered as a perturbation. 
In this case, we consider the full Hamiltonian as ${\mathcal H}_T={\mathcal H}_{\bf k} + H_{\rm{SO}}$ with ${\mathcal H}_{\bf k}$ corresponding to Eq.~\eqref{eq:hamorbitalRashba} of the main text.  In this case, the effect of spin-orbit coupling can be taken into account by performing a Schrieffer-Wolff transformation and deducing the effective spin Hamiltonian for the  spin-one-half doublet with $p_z$ orbital character. Then the effective Hamiltonian 
\begin{equation}
{\mathcal H}_{\rm{spin}}({\bf k}) = \sum_{h} \frac{\langle \ell | H_{\rm SO} | h \rangle \langle h | {\mathcal H}_{\bf k} | \ell \rangle + \text{h.c.}}{E_\ell - E_h} 
%%+ \langle \ell | H_{\rm SOC} | \ell \rangle.
\end{equation}
where the $|\ell\rangle= p_x \ket{\uparrow, \downarrow}$ while $|h\rangle$ is the quartet of degenerate states related to the $p_x,p_y$ orbitals. 
The final form of the effective spin-Hamiltonian is

\[
{\mathcal H}_{\rm{spin}}({\bf k})  =
\frac{2\lambda_{SOC} \alpha_{OR}}{\Delta}\,(k_x \sigma_y - k_y \sigma_x)+ \dfrac{\hbar^2 k^2}{2 m^{\star}} \sigma_0
\]
with $\sigma_0$ the identity matrix.
\begin{acknowledgments}
    M.C. acknowledges partial support by Italian Ministry of University and Research (MUR) PRIN 2022
under the Grant No. 2022LP5K7 (BEAT) and from PNRR MUR project PE0000023-NQSTI. M.T.M. and C.O.  acknowledge partial support  from PNRR MUR Project
No. PE0000023-NQSTI (TOPQIN).
\end{acknowledgments}

%\item[Author contributions:] 
{\bf Author contributions:} All authors contributed extensively to the work presented in the paper. 

%\item[Competing interests:]
{\bf Competing interests:} The authors declare that there are no competing financial or non-financial interests. 
%\end{addendum} 

\end{document}

\title{Supplementary Information for: \\``Curvature-induced nonlinear anomalous Hall effect in thin magnetic shells"} 
%Nonlinear anomalous Hall effect in curved magnetic membranes

\author{Maria Teresa Mercaldo}
\affiliation{Dipartimento di Fisica ``E. R. Caianiello", Universit\`a di Salerno, IT-84084 Fisciano (SA), Italy}

\author{Mario Cuoco}
\affiliation{CNR-SPIN, c/o Universit\`a di Salerno, IT-84084 Fisciano (SA), Italy}

\author{Carmine Ortix}
\affiliation{Dipartimento di Fisica ``E. R. Caianiello", Universit\`a di Salerno, IT-84084 Fisciano (SA), Italy}
\affiliation{CNR-SPIN, c/o Universit\`a di Salerno, IT-84084 Fisciano (SA), Italy}

\maketitle

\section{Quantum metric contribution to non linear conductivity}

In this note we discuss in detail the behavior of the nonlinear conductivity contributions driven by the quantum metric in a noncentrosymmetric two-dimensional ferromagnet described by Rashba spin–orbit interaction and a uniform magnetization.
The second order intrinsic % $(\tau^0)$ 
nonlinear conductivity is governed by the band normalized quantum metric (BNQM) dipoles $G^n_{ab}$. Indeed we have
\begin{eqnarray}
 \sigma_{a;bc}^{QM} &=& -\frac{e^3}{\hbar}\sum_{\substack{n \\  \text{(occupied)}}}\int \frac{d^2 k}{(2\pi)^2} \Lambda^{n}_{a;bc}(\bk)f_n(\bk),
\end{eqnarray}
where  $f_n(\bk)=f(\eps_n(\bk))$ is the equilibrium Fermi-Dirac distribution function of the $n$-band  and $a,b,c$ are cartesian coordinates and the %{\it kernel} 
functions $\Lambda_{a;bc}$ in the integral are expressed in term of first derivatives of the  BNQM $G^n_{ab}$:
\begin{eqnarray}
    \Lambda^{n}_{a;bc}(\bk) &=& \left[\frac32 \partial_a G_n^{bc} - \frac12(\partial_{b} G^n_{ca}+\partial_{c} G^n_{ab}) \right],
\end{eqnarray}
where $\partial_a \equiv \partial /\partial k_a$. 

In the following, we consider separately the model Hamiltonians $\mathcal{H}_{1,2}^{\rm spin}$, introduced in Eqs.(5)-(6) of the main text, which capture different symmetry properties of the system.

\subsection{${\mathcal C}_2 \Theta$ symmetric systems}

\begin{figure*}[t]
    \begin{center}        
    {\large (a)}\includegraphics[width=0.3\textwidth]{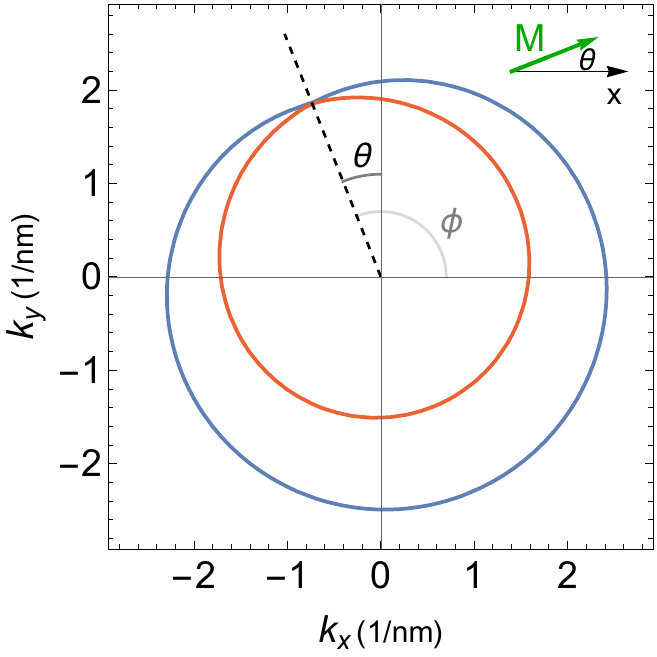}   \hspace{0.5cm}   
    \large{(b)}\includegraphics[width=0.3\textwidth]{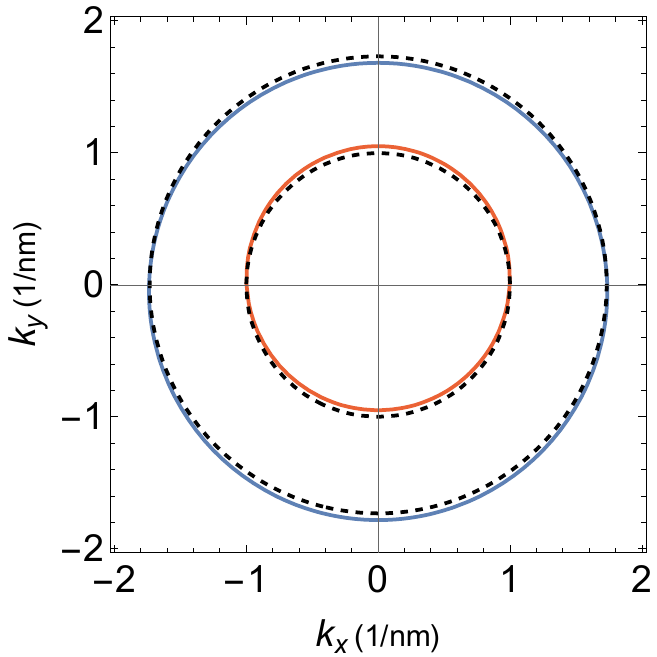} 
    \caption{(a) Fermi contours for $\alpha_R=0.2 $eV\r{A}, $M=40 $meV and $E_F=E_c=160 $meV. The band crossing occurs at $k_c= M/\alpha_R= 2 (1/$nm) and $\phi=\pi/2+\theta$. (b) Fermi lines for  $\alpha_R=0.04 eV$\r{A}, $M=40 $meV, $E_F=80 $meV and $\theta=0$. For this values of the parameters $E_c=4 $eV. The dashed lines are the circular lines obtained  for $\alpha_R=0$, which can be used for obtaining analytical approximations in when $M\gg \alpha_R k_F$.
     \label{fig:S1}}
    \end{center}
\end{figure*}

We start by analyzing  the transverse components of the conductivity tensor for the model Hamiltonian in Eq.(5) of the main text, i.e.
\begin{equation}
{\mathcal H}_1^{\rm{spin}}({\bf k})  =
\alpha_R \,(k_x \sigma_y - k_y \sigma_x)+ \dfrac{\hbar^2 k^2}{2 m^{\star}} \sigma_0 + {\bf M} \cdot \boldsymbol{\sigma}
\label{eqS:hamspinRashba}
\end{equation}
%when $\lambda=0$, so for our model
 For this model, and limiting ourselves to transverse transport, we have:
\begin{eqnarray}
    \Lambda^{\pm}_{x;yy}(\bk) &=& \frac32 \partial_x G^\pm_{yy} - \partial_{y}G^\pm_{xy} = \mp \frac{\alpha_R^3(\alpha_R k_x+M\sin\theta)[7(\alpha_R k_x+M\sin\theta)^2+2(\alpha_R k_y-M\cos\theta)^2]}{8 [(\alpha_R k_x+M\sin\theta)^2+(\alpha_R k_y-M\cos\theta)^2]^{7/2}}\\
    \Lambda^{\pm}_{y;xx}(\bk) &=& \frac32 \partial_y G^\pm_{xx}- \partial_{x}G^\pm_{xy} = \mp \frac{\alpha_R^3(\alpha_R k_y-M\cos\theta)[2(\alpha_R k_x+M\sin\theta)^2+7(\alpha_R k_y-M\cos\theta)^2}{8 [(\alpha_R k_x+M\sin\theta)^2+(\alpha_R k_y-M\cos\theta)^2]^{7/2}}
\end{eqnarray}
%%%%%%%%%%%%%%%%%%%%%%%
The energy bands of the spin Hamiltonian \eqref{eqS:hamspinRashba} are shown in Fig.3a of main paper, while in Fig.S\ref{fig:S1} we show the Fermi contours for a representative set of parameters.  
%We can distinguish two regimes for values of Fermi energy $E_F$ which are above or below  the energy at the band-crossing:
There is a %symmetry protected
band crossing at energy:
\begin{equation}
\label{eq:EC}
    E_c=\frac{\hbar^2 M^2}{2 m^*\alpha_R^2}
\end{equation}
which  occurs at $(k_x,k_y)=(-M\sin\theta /\alpha_R, M\cos\theta /\alpha_R) $ or, 
  at $k=M/\alpha_R$  for $\phi=\pi/2+\theta$,
using polar coordinates $(k,\phi)$ (see Fig.~S~\ref{fig:S1}(a)).

We can distinguish two regimes in the properties of the Fermi contours, separated by the band-crossing energy. For $E_F> E_c$, the Fermi lines wind around the crossing point, whereas for $E_F<E_c$ they encircle regions away from this point. Remarkably, in the former case it is possible to obtain an exact analytical expression for the nonlinear conductivity tensor $\sigma^{QM}_{a,bb}$.  In contrast, for  $E_F < E_c$,  analytical  expressions can only derived in the approximation  $M\gg \alpha_R k_F$, which is nevertheless the parameter regime relevant for the present study.

\begin{figure*}[tb]
    \begin{center}        
    \includegraphics[width=0.7\textwidth]{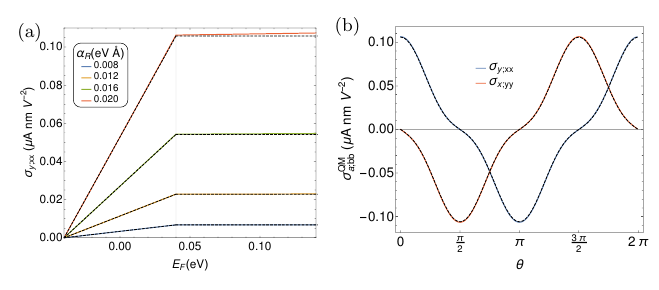}      
    \caption{(a)-(b) Comparison between analytical and numerical results. The analytical functions are drawn with black dashed lines. In (a) we show the behavior of $\sigma_{y;xx}^{QM}$ as a function of $E_F$. Here the analytical function is  given by Eqs.\eqref{eq:analytic1}-\eqref{eq:analytic2} and $E_L\simeq M =0.04 $eV. In (b) we show he comparison with the analytical expressions \eqref{eq:angYXX}-\eqref{eq:angXYY} of the angular behavior of the conductivity tensor components $\sigma^{QM}_{a;bb}(\theta)$ for $M=40$meV, $\alpha_R=0.02 eV$\r{A} and $E_F=60$mev. 
     \label{fig:S2}}
    \end{center}
\end{figure*}

\bigskip 
{\bf Regime 1:} $E_F < E_c$. % with $M\gg \alpha_R k_F$.
\\
We start our analysis with the case in which the ratio $\alpha_R k_F /M \ll 1 $. 
%can be considered a small parameter. 
Thus the Fermi contours can be approximated with circular lines of equations 
$$\frac{\hbar^2k^2}{2 m^*}\pm M =E_F,$$
which is a zero order approximation in $\alpha_R k_F/M$ (see also Fig.~S~\ref{fig:S1}(b)).
We next expand the functions $\Lambda_{a,bb}$ to leading order in $\alpha_R k_F/M$, obtaining:
\begin{eqnarray}
     \Lambda^{-}_{y;xx}(\bk) &=&  -\frac{\alpha_R^3}{16 M^4}\left[ (9+5 \cos(2\theta)) \cos\theta +O\left(\frac{\alpha_R k}{M}\right)\right]\\
     \Lambda^{-}_{x;yy}(\bk) &=&  -\frac{\alpha_R^3}{16 M^4}\left[ (9-5 \cos(2\theta)) \sin\theta+O\left(\frac{\alpha_R k}{M}\right)\right]
\end{eqnarray}
Let's start analyzing the case $\theta=0$.  To evaluate $\sigma_{y,xx}^{QM}$ we have to distinguish wether the Fermi energy is below or above the energy $E_L$ associated to the Lifshitz transition (i.e. we distinguish  situations in which only the first band is occupied and that where both bands are occupied). Indeed, in the first case we have to integrate  $\Lambda^{(-)}_{y,xx}(\bk)$ on a circle up to $k_{F_-}=\sqrt{(M+E_F)/c_2}$ (where $c_2 \equiv \hbar^2/(2 m^*)$), while in the latter case (considering that $\Lambda^{(+)}=-\Lambda^{(-)}$, the integration is in the Fermi annulus between $k_{F_+}=\sqrt{(E_F-M)/c_2}$ and $k_{F_-}$. %\BLU{
The Lifshitz transition occurs exaclty at the energy of the band-crossing $E_L=E_c$  for $M \leq \alpha_R^2/(2 c_2)$, while for $M > \alpha_R^2/(2 c_2)$ -- which is the situation we are studying -- we have   $E_L=M -\alpha_R^2/4 c_2 $. Thus, in the present approximation we can consider $E_L \simeq M$. %}
Notice that at this order of approximation, the functions $\Lambda_{a,bb}$ do not depend on $\bk$, thus the momentum integration will give simply the area enclosed in the Fermi lines: for $E_F<E_L$  the result will be proportional to $k_{F_+}^2 = (M+E_F)/c_2$, while for $E_F> E_L$ we will have the area of the annulus (i.e $\propto (k_{F_+}^2- k_{F_-}^2)=2M/c_2$). Specifically, we find 
\begin{eqnarray}
\label{eq:analytic1}
    \sigma_{y;xx}^{QM}&=& \frac{e^2}{\hbar} \frac{7}{32\pi} \frac{\alpha_R^3}{M^4}\left(\frac{2m^*}{\hbar^2}\right) (M+E_F)\quad \quad
    \text{ for }E_F < E_L
    %\text{if only one band is occupied}
    \\
\label{eq:analytic2}
      \sigma_{y;xx}^{QM}&=& \frac{e^2}{\hbar} \frac{7}{16\pi} \frac{\alpha_R^3}{M^3}\left(\frac{2m^*}{\hbar^2}\right)\qquad \qquad \qquad \;\text{ for }E_F > E_L.
      %\text{if both bands are occupied}
\end{eqnarray}
The comparison between the numerical results and the analytical approximation is shown in Fig.~S~\ref{fig:S2}. In panel (a), for the smallest values of $\alpha_R$, we clearly observe linear behavior  while varying $E_F$  up to the Lifshitz transition, followed by a constant regime. Conversely, for the highest value of $\alpha_R$, deviations from this {\it zero order} approximation become visible. While the analytical model could be improved by including second-order corrections in $\alpha_R k_F / M$ (noting that first-order corrections vanish), such an expansion is not relevant to understand the origin of the cusp. %lies beyond the scope of the present analysis.

For $\theta\neq 0$, considering 
$E_F> E_L$, we obtain:
\begin{eqnarray}
\label{eq:angYXX}
    \sigma_{y;xx}^{QM} &=& +\frac{e^3}{\hbar} \frac1{32\pi} \frac{\alpha_R^3}{M^3}\left(\frac{2 m^*}{\hbar}\right)\left(9+5 \cos(2\theta)\right) \cos\theta \\
\label{eq:angXYY}
    \sigma_{x;yy}^{QM} &=& -\frac{e^3}{\hbar} \frac{1}{32\pi} \frac{\alpha_R^3}{M^3}\left(\frac{2 m^*}{\hbar}\right)\left(9-5 \cos(2\theta)\right) \sin\theta .
\end{eqnarray}
The comparison of the angular behavior is shown in Fig.~S~\ref{fig:S2}(b).

%\medskip 
%\begin{figure*}[h]
%    \begin{center}        
%    \includegraphics[width=0.34\textwidth]{forCURVED_FM/fig3QM_bigAr.pdf}      
%    \caption{ $\sigma_{y,xx}^{QM}$ as a function of $E_F$ for $M=40 $meV. There is cusp in the behavior of the conductivity which occurs at the Lifshitz transition. Here we clearly see that this point depends also $\alpha_R$. For $\alpha_R=0.02 $eV\r{A}, the energy of the crossing point is at $E_c=0.16$eV. 
 %    \label{fig:BigAR}}
 %   \end{center}
%\end{figure*}

\begin{figure*}[h]
\centering
\begin{minipage}{0.4\textwidth}
    \includegraphics[width=0.9\textwidth]{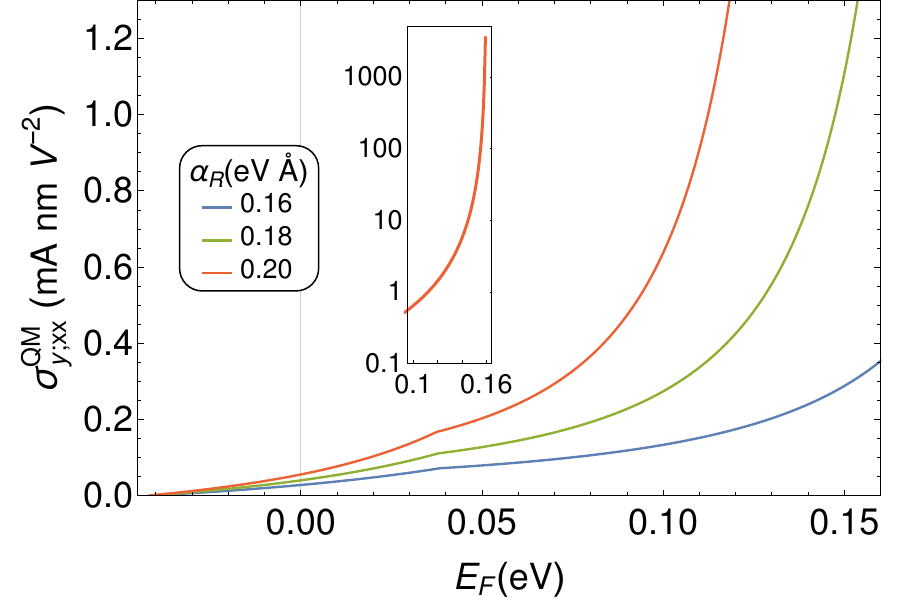}
\end{minipage}
\begin{minipage}{0.45\textwidth}
    \caption{ $\sigma_{y;xx}^{QM}$ as a function of $E_F$ for $M=40 $meV. There is cusp in the behavior of the conductivity which occurs at the Lifshitz transition. Here we clearly see that this point depends also $\alpha_R$. For $\alpha_R=0.02 $eV\r{A}, the energy of the crossing point is at $E_c=0.16$eV. 
     \label{fig:S3}}
\end{minipage}
\end{figure*}

For completeness, in  Fig.~S~\ref{fig:S3} we show the behavior of the intrinsic contribution to conductivity $\sigma_{y;xx}^{QM}$ for $E_F <E_c$  for larger values of the Rashba parameter $\alpha_R$, where the expansion in $\alpha_R k_F/M$ is no longer valid.
When $\lambda=0$, we see a divergence for $E_F\to E_c$.

\bigskip 
{\bf Regime 2:} $E_F > E_c$
\\
For this analysis we follow the strategy discussed in the Supplementary Information of Ref.\cite{Sala_24}. Indeed, in this case it is convenient to introduce shifted momenta
$p_x=k_x+\frac{M}{\alpha_R}\sin\theta$ and  $p_y=k_y -\frac{M}{\alpha_R}\cos\theta$, so that the Hamiltonian $\bd$-vector simplifies to $\bd=(-\alpha_R p_y,\alpha_R p_x,0)$. The advantage is that the BNQM-components, and hence their dipoles, do not depend on $M$. In particular we have:
\begin{eqnarray}
    G^\pm_{xx}=\pm \frac{p_y^2}{4|\alpha_R|p^{5}},\quad 
    G^\pm_{xy}=\mp \frac{p_x p_y}{4|\alpha_R|p^{5}},\quad
    G^\pm_{yy}=\pm \frac{p_x^2}{4|\alpha_R|p^{5}}
\end{eqnarray}
with  $p^2=p_x^2+p_y^2$, and for the $\Lambda_{a;bb}$-functions:
\begin{eqnarray}
%    \Lambda^{\pm}_{x;xx} &=& \mp \frac{5 p_x p_y^2}{8|\alpha_R|p^7}, \qquad     \quad \Lambda^{\pm}_{y;yy} = \mp \frac{5 p_x^2 p_y}{8|\alpha_R|p^7} \\
    \Lambda^{\pm}_{y;xx} &=& \mp \frac{p_y(2 p_x^2+7 p_y^2)}{8|\alpha_R|p^7}, \quad
    \Lambda^{\pm}_{x;yy} = \mp \frac{p_x(7p_x^2+2p_y^2)}{8|\alpha_R|p^7} .
\end{eqnarray}
The dependence on $M$ will appear in the conductivity tensor from the $M$-dependence  of the Fermi contours. 
Indeed, using shifted momenta, the energy reads
\begin{eqnarray}
\label{eq:Ep}
    E_\pm = \frac{\hbar^2}{2m^*}\left(p^2+\frac{M^2}{\alpha_R^2}+\frac{2 M p\sin(\phi-\theta)}{\alpha_R}\right)\pm|\alpha_R|p
\end{eqnarray}
and the degeneracy point will be always at the origin $(p_x=0,p_y=0)$. %The value of the energy at the crossing is:
%\begin{equation}
%\label{eq:EC}
%    E_c=\frac{\hbar^2 M^2}{2 m^*\alpha_R^2}
%\end{equation}

\begin{figure*}[t]
    \begin{center}        
    \includegraphics[width=0.97\textwidth]{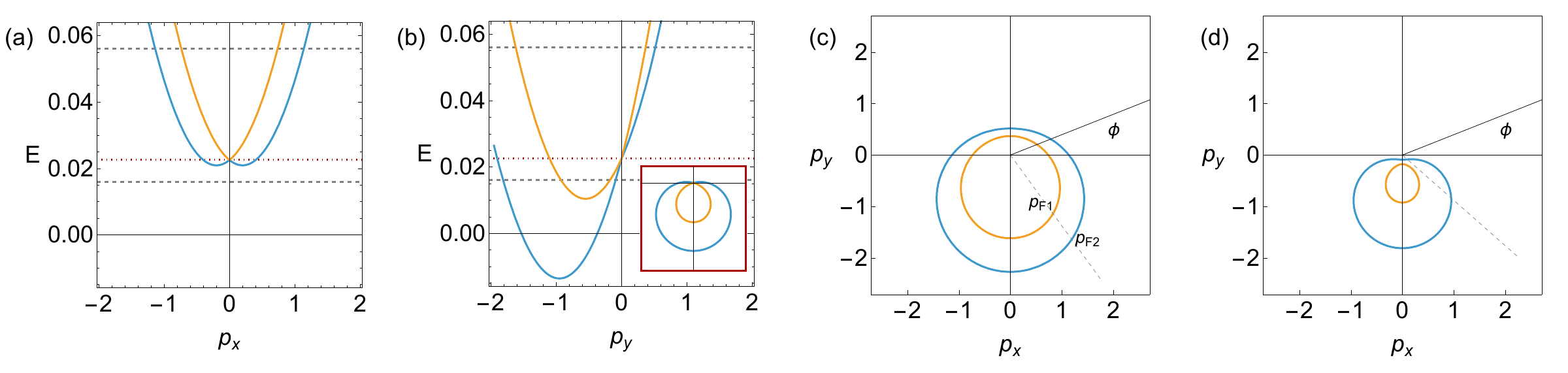}      
    \caption{(a)-(b) Energy bands for a representative set of parameters (specifically: $M=12 $meV and $\alpha_R=0.16 $eV\r{A}), as a function of $p_x$, with $p_y=0$ in (a) and as a function of $p_y$ for $p_x=0$ in (b). In the plots energy are in eV and momenta in (1/nm). The red dotted line shows the energy of the bands-crossing $E_c$, which in shifted momentum space occurs at the origin $(p_x=0,p_y=0)$. In the inset in (b) we display the Fermi lines at $E=E_{c}$. (c)-(d) Fermi lines %for the same set of parameters of (a)-(b) and 
    for two different values of Fermi energy, which are marked by the gray dashed lines in panels (a)-(b). We notice that in (c), which corresponds {\it regime 2} ($E_F> E_c$), for each value of $\phi\in [0,2\pi]$ we will have one solution $p_{Fi}$ for each Fermi contour, while in (d), corresponding to {\it regime 1}, the Fermi lines are not winding around the origin. Thus, there are values of the angle $\phi$ for which there are two solutions and values for which there are none. Indeed, as we can clearly see in panel (a), the lowest energy gray line does not intercept the energy bands along the $p_y=0$ line.
     \label{fig:S4}}
    \end{center}
\end{figure*}

In the present regime, $E_F>E_c$,   we have that both Fermi lines wind around the origin $(p_x=0,p_y=0)$ of the {\it shifted}-momentum space, as shown in Fig.~S~\ref{fig:S4}(c). So, for any $\phi$ we find a value for the Fermi wave-vector $p_{F(1,2)}(\phi)$. Instead, for $E_F<E_c$, i.e. the situation discussed above, using shifted momenta we see that for certain values of the angle $\phi$ there are two solutions or no solutions at all (see Fig.~S~\ref{fig:S4}(d)). 
In the present regime  both bands are occupied, thus we have  
\begin{eqnarray}
    \sigma_{a;bb}^{QM}&=&-\frac{e^3}{\hbar}\sum_{\substack{n \\  \text{(occupied)}}}\int \frac{d^2 k}{(2\pi)^2} \Lambda^{n}_{a;bb}(\bk)f_{0}(\eps_n(\bk)) =\nonumber \\
    &=&-\frac{e^3}{\hbar}\frac{1}{4\pi^2}\int d p_x d p_y \Lambda^{-}_{a;bb}(\bp)[f_{0}(\eps_-(\bp)) -f_{0}(\eps_+(\bp)) ] %=\\
%    &=& -\frac{e^3}{\hbar}\frac{1}{4\pi^2}\int_0^{2\pi} d\phi \int_{p_1(\phi)}^{p_2(\phi)} p\;dp \Lambda^{(-)}_{a,bb}(p,\phi)
\label{eq:sigmaGen}
\end{eqnarray}

%For completeness we consider here also the longitudinal cases.
The $\Lambda_{a,bb}$ functions in polar coordinates $(p,\phi)$ reads:
\begin{eqnarray*}
\begin{array}{l  l}
%\Lambda_{x,xx}=\ds\frac{5\sin^2\phi\cos\phi}{8|\alpha_R| p^4},\qquad &\Lambda_{y,yy}=\ds\frac{5\cos^2\phi\sin\phi}{8|\alpha_R| p^4} \\
\Lambda^-_{y;xx}=\ds\frac{2\sin\phi\cos^2\phi+7\sin^3\phi}{8|\alpha_R| p^4},\qquad &\Lambda^-_{x;yy}=\ds\frac{7\cos^3\phi+2\cos\phi\sin^2\phi}{8|\alpha_R| p^4} 
\end{array}
\end{eqnarray*}
so we can write in general $\Lambda^-_{a;bb}=f_{a;bb}(\phi)/8|\alpha_R|p^4$. The angular functions in both %all 
cases have the property $f_{a;bb}(\phi+\pi)=-f_{a;bb}(\phi)$, which will be used later in the explicit derivation.

To proceed we need the explicit expressions for $p_{F(1,2)}(\phi)$, solutions of the equations $E_\pm=E_F$.
To shorten the notation we introduce the coefficients 
\begin{equation}
    B_{1,2} =  \frac{M}{\alpha_R}\sin(\phi-\theta) \pm\frac{m^*|\alpha_R|}{\hbar^2},\quad  C= \frac{2m^*E_F}{\hbar^2}-\frac{M^2}{\alpha_R^2},
    \label{eq:BeC}
\end{equation} 
so the equations simplifies to $p^2+2 B_{1,2} p-C =0 $.
Notice that the conditions $E_F> E_c$ corresponds to $C>0$, hence, to get only one positive solution for each equation, we have to take only  the solution with positive sign: $p_{F(1,2)}(\phi)=-B_{(1,2)} + \sqrt{B_{(1,2)}^2 +C}$.

So, continuing from eq.\eqref{eq:sigmaGen}, we have
\begin{eqnarray}
    \sigma_{a;bb}^{(0)}&=& -\frac{e^3}{\hbar}\frac{1}{4\pi^2}\int_0^{2\pi} d\phi \int_{p_1(\phi)}^{p_2(\phi)} p\;dp\; \Lambda^{-}_{a;bb}(p,\phi) = \nonumber \\
     &=&-\frac{e^3}{\hbar}\frac{1}{32|\alpha_R|\pi^2}\int_0^{2\pi} d\phi \;f_{a;bb}(\phi) \int_{p_1(\phi)}^{p_2(\phi)} dp \frac1{p^3} =\nonumber \\
     &=&-\frac{e^3}{\hbar}\frac{1}{64 |\alpha_R|\pi^2}\int_0^{2\pi} d\phi\;f_{a;bb}(\phi) \left[\frac{1}{p^2_1(\phi)}-\frac{1}{p^2_2(\phi)}\right] \label{eq:sigmaint}
\end{eqnarray}
Now is it possible to show that $\left[\frac{1}{p^2_1(\phi)}-\frac{1}{p^2_2(\phi)}\right]$
 can be written as the sum of two functions %,  $g_{1}(\phi)$ and $g_2(\phi)$, 
 that behave differently while $\phi\in[0,2\pi]$:
\begin{eqnarray*}
    \left[\frac{1}{p^2_1(\phi)}-\frac{1}{p^2_2(\phi)}\right] &=& {\frac{1}{(-B_1+\sqrt{B_1^2+C})^2}- \frac{1}{(-B_2+\sqrt{B_2^2+C})^2} = }   \\
 %   &=&\BLU{\frac{(B_1+\sqrt{B_1^2+C})^2}{C^2}-\frac{(B_2+\sqrt{B_2^2+C})^2}{C^2} =}\\
    &=&{{\frac{2(B_1^2-B_2^2)}{C^2}} + {\frac{2B_1\sqrt{B_1^2+C}-2B_2\sqrt{B_2^2+C}}{C^2}}
=} g_1(\phi) + g_2(\phi)
\end{eqnarray*}
where
\begin{eqnarray}
    g_1(\phi) &=&\frac{2(B_1^2-B_2^2)}{C^2}= \frac{2 \hbar^2 M |\alpha_R|}{m^* \alpha_R } \frac{\sin(\phi-\theta)}{\left[E_F-\frac{M^2\hbar^2}{2 m^* \alpha_R^2}\right]^2} \\
    g_2(\phi) &=& {\frac{2B_1\sqrt{B_1^2+C}-2B_2\sqrt{B_2^2+C}}{C^2}}
\end{eqnarray}
Notice that for $\phi\to\phi+\pi$ we have $B_1\to -B_2$, so that $g_2(\phi+\pi)=g_2(\phi)$, while $g_1(\phi+\pi)=-g_1(\phi)$. Thus, reminding the behavior of $f_{a,bb}(\phi)$ it results that only $g_1(\phi)$ will give non-vanishing contribution to the angular integration. 
Inserting the expression of $g_1$ in \eqref{eq:sigmaint}, we find
\begin{eqnarray}
    \sigma_{a;bb}^{QM}&=& -\frac{e^3}{\hbar}\frac{1}{64|\alpha_R|\pi^2}\int_0^{2\pi} d\phi f_{a;bb}(\phi)g_1(\phi) = \nonumber \\
    &=&-\frac{e^3\hbar}{32\pi^2}\frac{M}{ m^* \alpha_R }
    \frac{1}{\left[E_F-\frac{M^2\hbar^2}{2 m^* \alpha_R^2}\right]^2}\int_0^{2\pi} f_{a;bb}(\phi)\sin(\phi-\theta)
\end{eqnarray}
Finally, we obtain the exact analytical expression of the transverse components of the conductivity tensor  in the regime $E_F>E_{c}$:
\begin{eqnarray}
\label{eq:exact}
 \sigma_{y;xx}^{QM}&=& 
 %-\frac{e^3\hbar}{32\pi^2}\frac{M}{ m^* \alpha_R }    \frac{1}{\left[E_F-\frac{M^2\hbar^2}{2 m^* \alpha_R^2}\right]^2}\int_0^{2\pi} [2\sin\phi\cos^2\phi +7\sin^3\phi]\sin(\phi-\theta) \nonumber \\     &=&
 \frac{23}{128 \pi} \,\frac{ \,e^3 \hbar  }{m^* \alpha_R } \frac{M\,\cos\theta}{\left[E_F-%\frac{M^2\hbar^2}{2 m^* \alpha_R^2}
 E_c\right]^2} \\
 \sigma_{x;yy}^{QM}&=&- \frac{23}{128 \pi} \,\frac{ \,e^3 \hbar  }{m^* \alpha_R } \frac{M\,\sin\theta}{\left[E_F-
 %\frac{M^2\hbar^2}{2 m^* \alpha_R^2}
 E_c\right]^2}\,.
\end{eqnarray}
%and for the longitudinal transport we get
%\begin{eqnarray}
% \sigma_{x,xx}^{(0)}&=& \frac{5}{128\pi} \,\frac{ \,e^3 \hbar  }{m^* \alpha_R } \frac{M\,\sin\theta}{\left[E_F-\frac{M^2\hbar^2}{2 m^* \alpha_R^2}\right]^2} \\
% \sigma_{y,yy}^{(0)}&=&- \frac{5}{128 \pi} \,\frac{ \,e^3 \hbar  }{m^* \alpha_R } \frac{M\,\cos\theta}{\left[E_F-\frac{M^2\hbar^2}{2 m^* \alpha_R^2}\right]^2}.
%\end{eqnarray}
Notice that in the present regime the two components of the non-linear conductivity tensor show a pure sinusoidal behavior in $\theta$, while for $E_F<E_c$ higher-order angular contributions are present.

\bigskip

\begin{figure*}[h]
\centering
\begin{minipage}{0.4\textwidth}
    \includegraphics[width=0.9\textwidth]{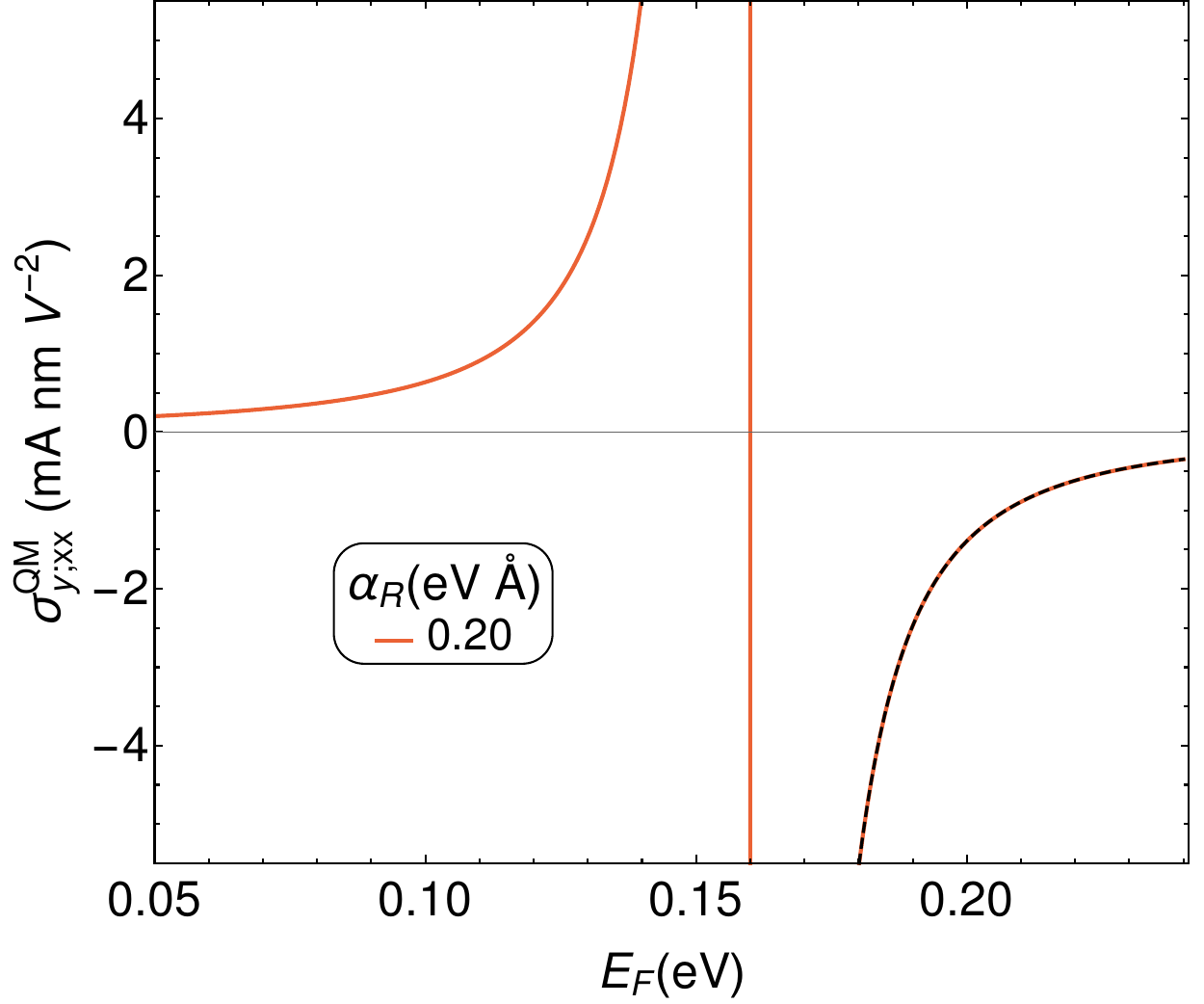}
\end{minipage}
\begin{minipage}{0.45\textwidth}
    \caption{ $\sigma_{y;xx}^{QM}$ as a function of $E_F$ for $M=40 $meV and $\alpha_R=0.2 $eV\r{A}. %which was shown in also in Fig.~S~\ref{fig:BigAR} up to $E_F=E_c=0.16 $eV. 
    The black dashed line is the analytical exact result Eq.\eqref{eq:exact}.
     \label{fig:S5}}
\end{minipage}
\end{figure*}

In Fig.~S~\ref{fig:S5} we show the behavior of the conductivity tensor in the two regimes: as already noted in Fig.~S~\ref{fig:S4}, it features a  divergence at $E_F=E_c$ with a sign change across this point. 

\subsection{Systems with broken rotational symmetries}

\begin{figure*}[h]
    \begin{center}        
    \includegraphics[width=1.\textwidth]{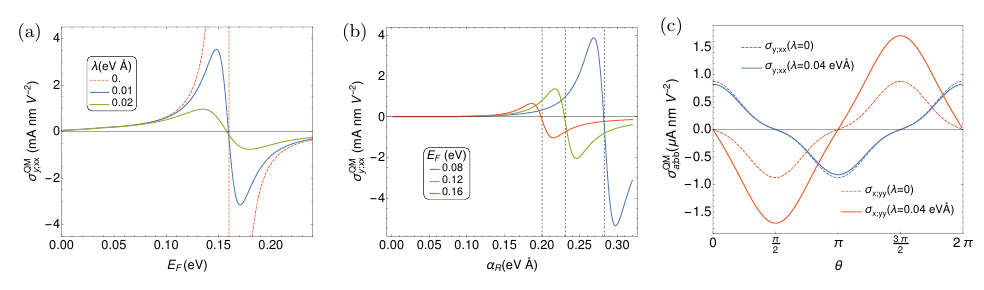}      
    \caption{In (a) we show the behavior of $\sigma_{y;xx}^{QM}$ as a function of $E_F$ for $\alpha_R=0.2 $eV\r{A}, $M=40$meV and different values of $\lambda$. The double-peak heights  decreases while increasing $\lambda$. In (b)  $\sigma_{y;xx}^{QM}$ as a function of $\alpha_R$ is displayed for three different values of $E_F$ (see legend). Here $M=40$meV and $\lambda=0.04 $eV\r{A}. The dashed vertical lines mark the values of   $\alpha_{R}$ corrsponding to $E_F=E_c$. (c) Comparison between the results for $\lambda=0$ and $\lambda= 0.04 $eV\r{A} of the angular behavior of $\sigma_{y;xx}^{QM}$ and  $\sigma_{x;yy}^{QM}$ for $M=40$meV, $\alpha_R=0.02 $eV\r{A} and $E_F=60$meV. 
     \label{fig:S6}}
    \end{center}
\end{figure*}

In this section we show how the additional term $\lambda k_y \sigma_z$ in the model Hamiltonian \eqref{eqS:hamspinRashba}, {which breaks rotation symmetry of the system},  modifies the intrinsic contribution to non linear conductivity. In Fig.~S~\ref{fig:S6}(a) we show the behavior of the non linear conductivity tensor  $\sigma_{y;xx}^{QM}$ as a function of $E_F$  for two values of $\lambda$. For comparison, we plot also the $\lambda =0$ case.
% -- which has been presented in Fig~S~\ref{fig:S5} -- 
The additional term in the Hamiltonian lifts the degeneracy of the bands (as shown in Fig.4a of main paper) and accordingly removes the divergent behavior of $\sigma^{QM}_{y;xx}$. Thus the non-linear conductivity tensor features a double peak and a sign change, which is pinned about $E_F=E_c$. Increasing the value of $\lambda$ the zero-point of the conductivity shifts towards smaller values of $E_F$.  

In Fig.~S~\ref{fig:S6}(b) we draw $\sigma_{y;xx}^{QM}$ vs. $\alpha_R$. The behavior resembles the one shown for the Berry curvature dipole (BCD) $D_y$ presented in Fig.4d of main paper. Notice, however, that $D_y$ does not contribute to $\sigma_{y;xx}$. Indeed we have:
%$\sigma^{(1)}_{x;yy}$  and to $\sigma^{(1)}_{y,;xy}=\sigma^{(1)}_{y;xy}$:
\begin{equation}
    \sigma^{(1)}_{x;yy} =\tau\frac{e^3 }{\hbar^2} D_y, \qquad  \sigma^{(1)}_{y;xx} =-\tau\frac{e^3 }{\hbar^2} D_x,%\qquad   \sigma_{y;xy}^{(1)} =\sigma_{y;yx}^{(1)}=-\tau\frac{e^3 }{\hbar^2} \frac{D_y}{2}.
\end{equation}
where the superscript $(1)$ refers to the power of the relaxation time $\tau$. Finally, in Fig.~S~\ref{fig:S6}(c) we show the angular dependence of $\sigma_{a;bb}^{QM}$, where $\theta$ is the angle between the magnetization and the $\hat{x}$ direction.  The additional term $\lambda k_y \sigma_z$ slightly affects $\sigma_{y;xx}^{QM}$ while enhances $\sigma_{x;yy}^{QM}$. Notice that, once rotational symmetry is broken,  the two conductivity components are no longer related by a $\pi/2$ shift, as occurs for $\lambda=0$.  

%\bibliography{biblionpjspintronics}